\documentclass[10pt]{elsart}

\usepackage{ifthen}
\usepackage{ifpdf}

\ifpdf
\usepackage{graphicx}
\usepackage{epstopdf}
\else
\usepackage{graphicx}
\usepackage{epsfig}
\fi

\usepackage{times}
\usepackage{float}
\usepackage{amsmath}
\usepackage{amssymb}
\usepackage{bm}
\usepackage{latexsym}

\begin{document}

\newcommand{\eexp}{\mbox{e}^}
\newcommand{\be}[1]{\begin{equation}\ifthenelse{#1=-1}{\nonumber}{\ifthenelse{#1=0}{}{\label{e#1}}}}
\newcommand{\ee}{\end{equation}} 
\newcommand{\cl}[1]{#1_{\tbox{cl}}}
\newcommand{\tbox}[1]{\mbox{\scriptsize#1}}
\newcommand{\ecl}{\delta E_{\tbox{cl}}}
\newcommand{\mbf}[1]{\mathbf{#1}}
\newcommand{\mpg}[2][1.0\hsize]{\begin{minipage}[b]{#1}{#2}\end{minipage}}
\newcommand{\mcal}[1]{\mathcal{#1}}
\newcommand{\cal}{\mathcal}
\newcommand{\szp}{\;.}
\newcommand{\szk}{\;,}
\newcommand{\dd}[1]{\:\mbox{d}#1}

\begin{frontmatter}

\title{Wavepacket Dynamics, Quantum Reversibility and Random Matrix Theory}

\author[a]{Moritz Hiller\corauthref{mh}},
\corauth[mh]{Corresponding author}
\ead{mhiller@chaos.gwdg.de}
\author[b]{Doron Cohen},
\author[a]{Theo Geisel},
\author[a]{Tsampikos Kottos}
 
\address[a]{Max-Planck-Institute for Dynamics and Self-Organization and \\
Department of Physics, University of G\"ottingen, Bunsenstra{\ss}e 10, D-37073 G\"ottingen}

\address[b]{Department of Physics, Ben-Gurion University, Beer-Sheva 84105, Israel }

\begin{abstract}
We introduce and analyze the physics of ``driving reversal" experiments. 
These are prototype wavepacket dynamics scenarios probing 
quantum irreversibility. Unlike the mostly hypothetical ``time reversal" concept,   
a ``driving reversal" scenario can be realized in a laboratory experiment, 
and is relevant to the theory of quantum dissipation.      
We study both the energy spreading and the survival probability in   
such experiments. We also introduce and study the "compensation time" 
(time of maximum return) in such a scenario. Extensive effort is devoted  
to figuring out the capability of either Linear Response Theory (LRT) 
or Random Matrix Theory (RMT) in order to describe specific features of the time evolution. 
We explain that RMT modeling leads to a strong non-perturbative response 
effect that differs from the semiclassical behavior. 
\end{abstract}

\end{frontmatter}

%%%%%%%%%%%%%%%%%%%%%%%%%%%%%%%%%%%%%%%%%%%%%%%%%%%%%%%%%%%%%%%%%%%%%%%
%%%%%%%%%%%%%%%%%%%%%%%%%%%%%%%%%%%%%%%%%%%%%%%%%%%%%%%%%%%%%%%%%%%%%%%
\section{Introduction}

In recent years there has been an increasing interest in understanding the theory of driven quantized 
chaotic systems \cite{W87,W88,WA95,BDK96,Wbook98,Wbook99,C00,CK00,WC02,KC03,CK03}. Driven systems 
are described by a Hamiltonian ${\cal H}(Q,P,x(t))$, where~$x(t)$ is a time dependent parameter 
and $(Q,P)$ are some generalized actions. Due to the time dependence of $x(t)$, the energy of 
the system is not a constant of motion. Rather the system makes "transitions" between energy 
levels, and therefore absorbs energy. This irreversible loss of energy is known as {\it dissipation}. 
To have a clear understanding of quantum dissipation we need a theory for the time evolution 
of the energy distribution.

Unfortunately, our understanding on quantum dynamics of chaotic systems is still quite limited. 
The majority of the existing quantum chaos literature concentrates on understanding the 
properties of eigenfunctions and eigenvalues. One of the main outcomes of these studies is 
the conjecture that Random Matrix Theory (RMT) modeling, initiated half a century ago by Wigner \cite{W55,W57}, 
can capture the {\it universal} aspects of quantum chaotic systems \cite{Sbook98,Hbook00}. Due 
to its large success RMT has become a major theoretical tool in quantum chaos studies 
\cite{Sbook98,Hbook00}, and it has found applications in both nuclear and mesoscopic physics 
(for a recent review see \cite{A00}). However, its applicability to quantum dynamics was left 
unexplored \cite{CIK00,KC01}.

%%%%%%%%%%%%%%%%%%%%
This paper extends our previous reports \cite{KC03,CIK00,KC01} on quantum {\em dynamics}, both
in detail and depth. Specifically, we analyze two dynamical schemes: The first is the so-called 
wavepacket dynamics associated with a rectangular pulse of strength $+\epsilon$ which is turned 
on for a specified duration; The second involves an additional pulse followed by the first one
which has a strength $-\epsilon$ and is of equal duration. We define this latter scheme as {\it 
driving reversal} scenario. We illuminate the direct relevance of our study with the studies of quantum 
irreversibility of energy spreading \cite{KC03} and consequently with quantum dissipation. We 
investigate the conditions under which maximum compensation is succeeded and define the notion 
of compensation (echo) time. To this end we rely both on numerical calculations performed for a 
chaotic system and on analytical considerations based on Linear Response Theory (LRT). The latter 
constitutes the leading theoretical framework for the analysis of driven systems and our study 
aims to clarify the limitations of LRT due to chaos. Our results are always compared with the 
outcomes of RMT modeling. We find that the RMT approach fails in general, to give the correct 
picture of wave-evolution. 
RMT can be trusted only to the extend that it gives trivial results that are implied by 
perturbation theory. Non-perturbative effects are sensitive to the underlying classical 
dynamics, and therefore the $\hbar\rightarrow 0$ behavior for effective RMT models is 
strikingly different from the correct semiclassical limit.

%%%%%%%%%%%%%%%%%%%%

The structure of this paper is as follows: In the next section we discuss the notion of
irreversibility which is related to driving reversal schemes and distinguish it from micro-reversibility
 which is associated with time reversal experiments. In Section \ref{sec:brm} 
we discuss the driving schemes that we are using and we introduce the various observables
that we will study in the rest of the paper. In Section \ref{sec:modelS}, the model systems 
are introduced and an analysis of the statistical properties of the eigenvalues
and the Hamiltonian matrix is presented. The Random Matrix Theory modeling is presented
in Subsection \ref{subsec:RMT}. In Section \ref{sec:regimes} we introduce the concept of parametric 
regimes and exhibit its applicability in the analysis of parametric evolution of eigenstates 
\cite{CK01}. Section \ref{sec:Meas_of_QI} extends the notion of regimes in dynamics and
presents the results of Linear Response Theory for the variance and the survival probability.
The Linear Response Theory (LRT) for the variance is analyzed in details in the following Subsection
\ref{sec:LRT}. In this subsection we also introduce the notion of restricted quantum-classical 
correspondence (QCC) and show that, as far as the second moment of the evolving wavepacket 
is concerned, both classical and quantum mechanical LRT coincides. In \ref{sec:LRTSP} we
present in detail the results of LRT for the survival probability for the two driving schemes 
that we analyze. The following Sections \ref{sec:WP-dyn} and \ref{sec:DR} contain 
the results of our numerical analysis together with a critical comparison with the theoretical 
predictions obtained via LRT. Specifically in Sect.\ref{sec:WP-dyn}, we present an analysis 
of wavepacket dynamics \cite{KC01} and expose the weakness of RMT strategy to describe 
wavepacket dynamics. In Sect.\ref{sec:DR} we study the evolution in the second half of the 
driving period and analyze the Quantum Irreversibility in energy spreading, where strong 
non-perturbative features are found for RMT models \cite{KC03}. Section 
\ref{sec:sum} summarizes our findings.

%%%%%%%%%%%%%%%%%%%%%%%%%%%%%%%%%%%%%%%%%%%%%%%%%%%%%%%%%%%%
%%%%%%%%%%%%%%%%%%%%%%%%%%%%%%%%%%%%%%%%%%%%%%%%%%%%%%%%%%%%
\section{Reversibility}

The dynamics of either a classical or a quantum mechanical system is generated by a Hamiltonian 
${\cal H}(Q,P;x(t))$ where $x=(X_1,X_2,X_3,...)$ is a set of parameters that can be controlled 
from the "outside". In principle $x$ stands for the infinite set of parameters that describe the 
electric and magnetic fields acting on the system. But in practice the experimentalist can control 
only few parameters. A prototype example is a gas of particles inside a container with a piston. 
Then $X_1$ may be the position of the piston, $X_2$ may be some imposed electric field, and $X_3$ 
may be some imposed magnetic field. Another example is electrons in a quantum dot where some 
of the parameters $X$ represent gate voltages.  

What do we mean by reversibility? Let us assume that the system 
evolves for some time. The evolution is described by 
\be{0}
U[x] = \mbox{Exp}\left( -{i\over \hbar}\int_0^t {\cal H}(x(t'))dt'\right)\szk
\ee
where ${\mbox{Exp}}$ stands for time ordered exponentiation. In the case of the archetype 
example of a container with gas particles, we assume that there is a piston (position $X$) 
that is translated outwards ($X^A(t)$ increasing). Then we would "undo" the evolution, by 
displacing the piston "inwards" ($X^B(t)$ decreasing). In such a case the complete evolution 
is described by $U[x] = U[x^B] U[x^A]$. If we get $U=1$ (up to a phase factor), then it means 
that it is possible to bring the system back to its original state. In this case we say that 
the process $U[x]$ is reversible.

In the strict adiabatic limit the above described process is indeed reversible. What about 
the non-adiabatic case? In order to have a well posed question we would like to distinguish 
below between "time reversal" and "driving reversal" schemes.

%%%%%%%%%%%%%%%%%%%%%%%%%%%%%%%%%%
\subsection{Time reversal scheme}
  
Obviously we are allowed to invent very complicated schemes in order 
to "undo" the evolution. The ultimate scheme 
(in the case of the above example) involves reversal of the velocities. 
Assume that this operation is represented by $U_T$,  
then  the reverse  evolution is described by 
\be{0}
U_{\tbox{reverse}} = U_T U[x^B] U_T \szk
\ee
where in $x_B(t)$ we have the time reversed piston displacement ($X(t)$) together with the 
sign of the magnetic field (if it exists) should be inverted. 
The question is whether $U_T$ can be realized.  
If we postulate that any unitary or anti-unitary 
transformation can be realized, 
then it follows trivially that any unitary evolution
is "micro-reversible". But when we talk about 
reversibility (rather than micro-reversibility) 
we allow control over a restricted set of parameters (fields).
Then the question is whether we can find a driving scheme, 
named $x^T$, such that
\be{0}
U_T = U[x^T]
\ \ \ \ \ \ \ \ \mbox{???}
\ee
With such restriction it is clear that in general 
the evolution is not reversible.

Recently it has been demonstrated in an actual 
experiment that the evolution of spin system 
(cluster with many interacting spins) can be reversed. 
Namely, the complete evolution was described 
by $U[x] = U[x^T] U[x^A] U[x^T] U[x^A]$,  
where $U[x^A]$ is generated by some Hamiltonian 
${\cal H}_A = {\cal H}_0 + \varepsilon {\cal W}$. 
The term ${\cal H}_0$ represents the interaction 
between the spins, while the term ${\cal W}$ 
represents some extra interactions. 
The unitary operation $U[x^T]$ is   
realized using NMR techniques, and its effect  
is to invert the signs of all the couplings. 
Namely $U[x^T] {\cal H}_0 U[x^T] = -{\cal H}_0$.
Hence the reversed evolution is described by 
\be{0}
U_{\tbox{reverse}} = 
\exp\left( -{i\over \hbar}t (-{\cal H}_0 + \varepsilon {\cal W})\right)\szk
\label{fidel}
\ee
which is the so-called Loschmidt Echo scenario. In principle we would like to have 
$\varepsilon =0$ so as to get $U=1$,  but in practice we have some un-controlled 
residual fields that influence the system, and therefore $\varepsilon \ne 0$. There 
is a huge amount of literature that discusses what happens in such scenario
\cite{PS02,BC02,JP01,JAB02,HKCG04}.

%%%%%%%%%%%%%%%%%%%%%%%%%%%%%%%%%%
\subsection{Driving reversal scheme}

The above described experiment is in fact exceptional. In most cases it is possible 
to invert the sign of only one part of the Hamiltonian, which is associated with the 
driving field. Namely, if for instance $U[x^A]$ is generated by ${\cal H}_A = H_0 + 
\varepsilon {\cal W}$, then we can realize 
\be{0}
U_{\tbox{reverse}} = 
\exp\left( -{i\over \hbar}t ({\cal H}_0 - \varepsilon {\cal W})\right)\szk
\ee
whereas  Eq.(\ref{fidel}) cannot be realized in general. 
We call such a typical scenario ``driving reversal" in order 
to distinguish it from ``time reversal" 
(Loschmidt Echo) scenario.

The study of ``driving reversal" is quite different from the study of ``Loschmidt Echo". 
A simple minded point of view is that the two problems are formally equivalent because 
we simply permute the roles of ${\cal H}_0$ and ${\cal W}$. In fact there is no symmetry 
here. The main part of the Hamiltonian has in general an unbounded spectrum with well 
defined density of states, while the perturbation ${\cal W}$ is assumed to be bounded. 
This difference completely changes the ``physics" of dynamics.

To conclude the above discussion we would like 
to emphasize that micro-reversibility is 
related to ``time reversal" experiment which 
in general cannot be realized, while the issue 
of reversibility is related to ``driving reversal",
which in principle can be realized.
Our distinction reflects the simple observation 
that not any unitary or anti-unitary operation 
can be realized.

%%%%%%%%%%%%%%%%%%%%%%%%%%%%%%%%%%%%%%%%%%%%%%%%%%%%%%%%%%%%%%
%%%%%%%%%%%%%%%%%%%%%%%%%%%%%%%%%%%%%%%%%%%%%%%%%%%%%%%%%%%%%%
\section{Object of the Study \label{sec:brm}}

In this paper we consider the issue of 
irreversibility for quantized chaotic systems.
We assume for simplicity one parameter driving. 
We further assume that the variation 
of $x(t)$ is small in the corresponding classical system so that the 
analysis can be carried out with a linearized Hamiltonian. 
Namely, 
\begin{equation}
{\cal H}(Q,P;x(t)) \approx {\cal H}_0 + \delta x(t) {\cal W}\szk
\label{eq:model-hamilt-class}
\end{equation}
where ${\cal H}_0 \equiv {\cal H}(Q,P;x(0))$ and $\delta x= x(t)-x(0)$. 
For latter purposes it is convenient to write the perturbation as 
\be{0}
\label{deltaxn1}
\delta x(t)= \varepsilon \times f(t)\szk
\ee
where $\varepsilon$ controls the "strength of the perturbation", 
while $f(t)$ is the scaled time dependence. 
Note that if $f(t)$ is a step function, 
then $\varepsilon$ is the "size" of the perturbation, 
while if $f(t)\propto t$  then $\varepsilon$ is the "rate" of the driving.  
In the representation of ${\cal H}_0$ we can write
\begin{equation}
{\cal H} = \bm{E} + \delta x(t) \bm{B}\szk
\label{eq:2DW_hamilton_matrx}
\end{equation}
where by convention the diagonal terms of $\bm{B}$ 
are absorbed into the diagonal matrix $\bm{E}$. 
From general considerations that we explain later 
it follows that $\bm{B}$ is a banded matrix 
that looks random. This motivates the study 
of an Effective Banded Random Matrix (EBRM) model, 
as well as its simplified version which is the 
standard Wigner Banded Random Matrix (WBRM) model.    
(See detailed definitions in the following).

In order to study the irreversibility for a given driving scenario, we have to introduce 
measures that quantify the departure from the 
initial state. We define a set of such measures 
in the following subsections.

%%%%%%%%%%%%%%%%%%%%%%%%%%%%%%%%%%%%%%%%%%%
\subsection{The evolving distribution $P_t(n|n_0)$}

Given the Hamiltonian ${\cal H}(Q,P;x)$, 
an initial preparation at state $|n_0\rangle$, 
and a driving scenario $x(t)$, 
it is most natural to analyze 
the evolution of the probability distribution 
\be{0}
\label{trprn1}
P_t(n|n_0) = |\langle n | U(t) | n_0 \rangle|^2\szp
\ee   
We always assume that $x(t)=x(0)$.

By convention we order the states by their energy. Hence we can regard ${\cal P}_t(n|n_0)$ 
as a function of $r=n-n_0$, and average over the initial preparation, so as to get a smooth 
distribution ${\cal P}_t(r)$.

The survival probability is defined as
\be{0}
\label{sptn1}
{\cal P}(t) = |\langle n_0 | U(t) | n_0 \rangle|^2 = P_t(n_0|n_0)\szk
\ee   
and the energy spreading is defined as
\be{0}
\label{ensptn1}
\delta E(t) = \sqrt{\sum_n P_t(n|n_0) (E_n-E_{n_0})^2}\szp
\ee   
These are the major measures for the characterization 
of the distribution. In later sections we would like 
to analyze their time evolution.

The physics of $\delta E(t)$ is very different from the physics 
of ${\cal P}(t)$ because the former is very sensitive to the 
tails of the distribution. Yet, the actual "width" of the distribution 
is not captured by any of these measures. A proper measure 
for the width can be defined as follows:
\be{0}
\label{cwtn1}
\delta E_{\tbox{core}}(t) \,\,=\,\, [n_{75\%} - n_{25\%}]\Delta\szk
\ee   
where $\Delta$ is the mean level spacing and $n_q$ is determined through the equation 
$\sum_n P_t(n|n_0) = q$. Namely it is the width of the main body of the distribution.
Still another characteristic of the distribution is 
the participation ratio $\delta n_{\tbox{IPR}}(t)$.
It gives the number of levels 
that are occupied at time $t$ 
by the distribution. 
The ratio $\delta n_{\tbox{IPR}} / (n_{75\%} - n_{25\%}) $ 
can be used as a measure for sparsity. We assume   
in this paper strongly chaotic systems, 
so sparsity is not an issue and 
$\delta n_{\tbox{IPR}} \sim  \delta E_{\tbox{core}} / \Delta $.

%%%%%%%%%%%%%%%%%%%%%%%%%%%%%%%%%%%%%%%%%%%
\subsection{The compensation time $t_r$}

\begin{figure}[t]
\centerline{\includegraphics[width=0.70\hsize,clip]{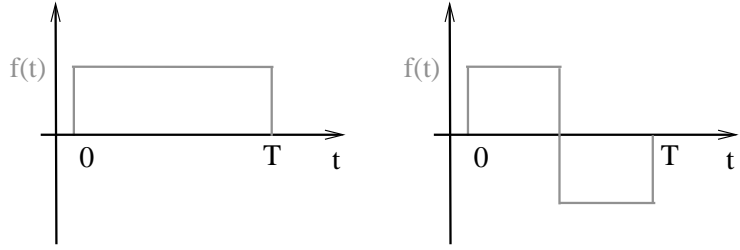}}
\caption{\label{cap:D:Shape-of-the}
Shape of the applied driving schemes $f(t)$; wavepacket dynamics (left panel) and
driving reversal scenario (right panel)}
\end{figure}

In this paper we consider two types of driving schemes.
Both driving schemes are presented schematically 
in Figure \ref{cap:D:Shape-of-the}. 

The first type of scheme is the {\it wavepacket dynamics} 
scenario for which the driving is turned-on at time 
$t=0$ and turned-off at a later time $t=T$. 

The second type of scenario that we 
investigate is what we call {\it driving reversal}. 
In this scenario the initial rectangular pulse 
is followed by a compensating pulse of equal duration.
The total period of the cycle is $T$.

In Figure \ref{cap:dE-2DW} we show representative results 
for the time evolution of $\delta E(t)$ in a wavepacket 
scenario, while in  Figure \ref{DRde} we show what happens 
in case of a driving reversal scenario. 
Corresponding plots for ${\cal P}(t)$ are presented 
in Figure \ref{DRpt}. We shall define the models and 
we shall discuss the details of these figures later on. 
At this stage we would like to motivate by inspection 
of these figures the definition of "compensation time".

We define the compensation time $t_r$, 
as the time after the driving reversal, 
when maximum compensation (maximum return) is observed. 
If it is determined by the maximum of the survival 
probability kernel ${\cal P}(t)$, then we denote it as $t_r^P$.  
If it is determined by the minimum of the 
energy spreading $\delta E(t)$  then we denote it as $t_r^E$.
It should be remembered that the theory 
of ${\cal P}(t)$ and $\delta E(t)$ is not the same, 
hence the distinction in the notation.
The time of maximum compensation is in general 
not $t_r=T$ but rather 
\begin{equation}
T/2<t_{r}<T\szp
\end{equation}
We emphasize this point because the notion of "echo", 
as used in the literature, seems to reflect 
a false assertion~\cite{HKCG04}.

%%%%  FIGURE %%%%%%%%%%%%%%%%%%%%%%%%%%
%%%% END FIGURE %%%%%%%%%%%%%%%%%%%%%%%

%%%%%%%%%%%%%%%%%%%%%%%%%%%%%%%%%%%%%%%%%%%%%%%%%%%%%%%%%%%%%%%%%%%%%%%%%%%%%%
%%%%%%%%%%%%%%%%%%%%%%%%%%%%%%%%%%%%%%%%%%%%%%%%%%%%%%%%%%%%%%%%%%%%%%%%%%%%%%

For the convenience of the reader we concentrate in the following table 
on the major notations in this paper: \\

{
\renewcommand{\baselinestretch}{0.85}
\small\normalsize

\begin{tabular}{|l|l|l|}
\hline

Notation &
explanation &
reference \\

\hline

${\cal H}(Q,P;x(t))$ &
classical linearized Hamiltonian &
Eq.(\ref{eq:model-hamilt-class}) \\
${\cal F}(t)$ &
generalized force  &
Eq.(\ref{eq:generalized-force}) \\

$C(\tau)$ &
correlation function &
Eq.(\ref{eq:autocorrelation})  \\

$\tau_{\tbox{cl}}$ &
correlation time  &
-- \\

$\tilde{C}(\omega)$ &
fluctuation spectrum  &
Eq.(\ref{powsp}) \\

\hline

${\cal H} = \bm{E}+\delta x  \bm{B}$ &
The Hamiltonian matrix  &
Eq.(\ref{eq:2DW_hamilton_matrx}) \\

2DW & 
the physical model system  &
Eq.(\ref{eq:2DW_hamiltonian}) \\

EBRM & 
the corresponding RMT model  &
-- \\

WBRM &
the Wigner RMT model &
--  \\

\hline

$\Delta$ &
mean level spacing  &
Eq.(\ref{mlsn1}) \\

$\Delta_b$ &
energy bandwidth  &
Eq.(\ref{blsn1}) \\

$\sigma$ &
RMS of near diagonal couplings  &
-- \\

$P_{\tbox{spacings}}(s)$ & 
energy spacing distribution  &
Eq.(\ref{eq:GOE}) \\
 
$P_{\tbox{couplings}}(q)$ &
distribution of couplings  &
Eq.(\ref{eq:wilkinson}) \\

\hline

$E_n(x)$ & 
eigen-energies of the Hamiltonian  &
-- \\

$E_n{-}E_m \approx r\Delta$ & 
estimated energy difference for $r=n-m$  &
-- \\

$P(n|m)$ & 
overlaps of eigenstates given a constant perturbation $\varepsilon$  &
Eq.(\ref{ef}) \\

$P(r)$ & 
smoothed version of $P(n|m)$   &
-- \\

$\Gamma(\delta x)$ & 
the number of levels that are mixed non-perturbatively &
-- \\

$\delta E_{\tbox{cl}} \propto \delta x$ & 
the classical width of the LDoS &
Eq.(\ref{e26}) \\

\hline

$\delta x = \varepsilon f(t)$ &
driving scheme    &
Eq.(\ref{deltaxn1}) \\

$T$ &
The period of the driving cycle (if applicable)  &
-- \\

$P_t(n|m)$ & 
the transition probability  &
Eq.(\ref{trprn1}) \\

$P_t(r)$ & 
smoothed version of $P_t(n|n_0)$   &
-- \\

$\mathcal{P}(t)$ & 
the survival probability $P_t(n_0|n_0)$ &
Eq.(\ref{sptn1}) \\

$p(t) = 1- \mathcal{P}(t)$ & 
total transition probability  &
Eq.(\ref{eq:FOPT-P}) \\

$\delta E(t)$ & 
energy spreading  &
Eq.(\ref{ensptn1}) \\

$\delta E_{\tbox{core}}(t)$ &
the "core" width of the distribution  &
Eq.(\ref{cwtn1}) \\

\hline

$t_r^P$ & 
compensation time for the survival probability  &
-- \\

$t_r^E$ & 
compensation time for the energy spreading  &
-- \\

$t_{\tbox{prt}}, \, t_{\tbox{sdn}}, \, t_{\tbox{erg}}$ &
various time scales in the dynamics  &
Eq.(\ref{eq:tprt},\ref{tergn1},\ref{e6}) \\

$\varepsilon_c, \,\, \varepsilon_{\tbox{prt}}$  & 
borders between regimes  &
Eq.(\ref{ecpn1},\ref{eprtn1}) \\

$P_{\tbox{FOPT}}, P_{\tbox{prt}}, P_{\tbox{sc}}$ &
various approximations to $P()$  &
Eq.(\ref{Pst1},\ref{Plor},\ref{Psc},\ref{Pcl}) \\

\hline

\end{tabular}
}

%%%%%%%%%%%%%%%%%%%%%%%%%%%%%%%%%%%%%%%%%%%%%%%%%%%%%%%%%%
\section{Modeling \label{sec:modelS}}

We are interested in quantized chaotic systems 
that have few degrees of freedom. 
The dynamical system used in our studies is 
the Pullen-Edmonds model \cite{PE81,M86}. 
It consists of two harmonic oscillators that 
are nonlinearly coupled. The corresponding Hamiltonian is
\begin{equation}
\mathcal{H}(Q,P;x)=
{1\over 2}\left(P_{1}^{2}+P_{2}^{2}+Q_{1}^{2}+Q_{2}^{2}\right)+
x  Q_{1}^{2}Q_{2}^{2}\szp
\label{eq:2DW_hamiltonian}
\end{equation}
The mass and the frequency of the harmonic oscillators are set to one. Without loss of 
generality we set $x(0)=x_0=1$. Later we shall consider classically small deformations 
(${\delta x \ll 1 }$) of the potential. One can regard this model (\ref{eq:2DW_hamiltonian}) 
as a description of a particle moving in a two dimensional well (2DW). The energy $E$ is 
the only dimensionless parameter of the classical motion. For high energies $E>5$ the motion 
of the Pullen-Edmonds model is ergodic. Specifically it was found that the measure of 
the chaotic component on the Poincar\'e section deviates from unity by no more than 
$10^{-3} $\cite{PE81,M86}.

\begin{figure}[t]
\centerline{
\includegraphics[width=0.49\hsize,clip]{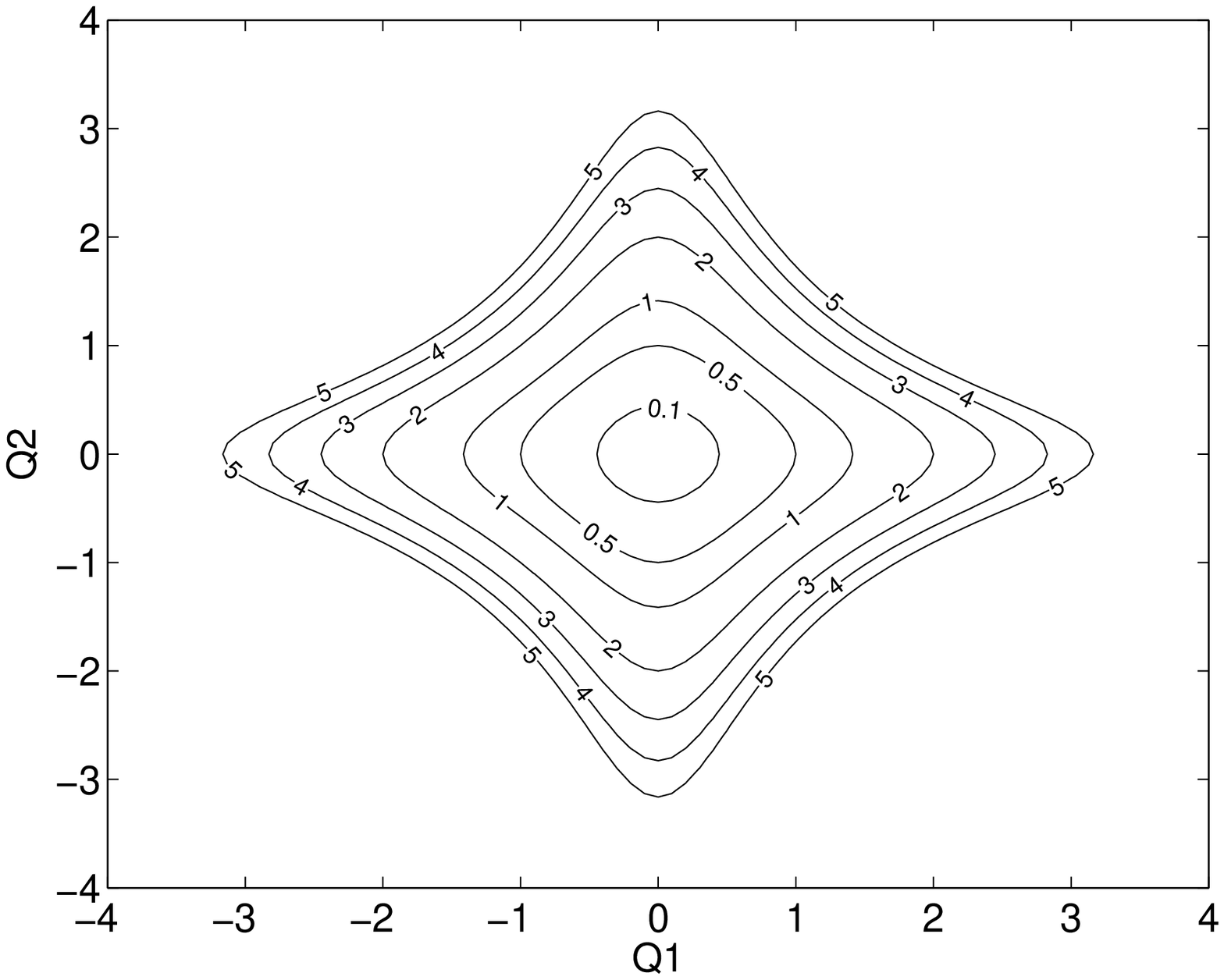}
\includegraphics[width=0.50\hsize,clip]{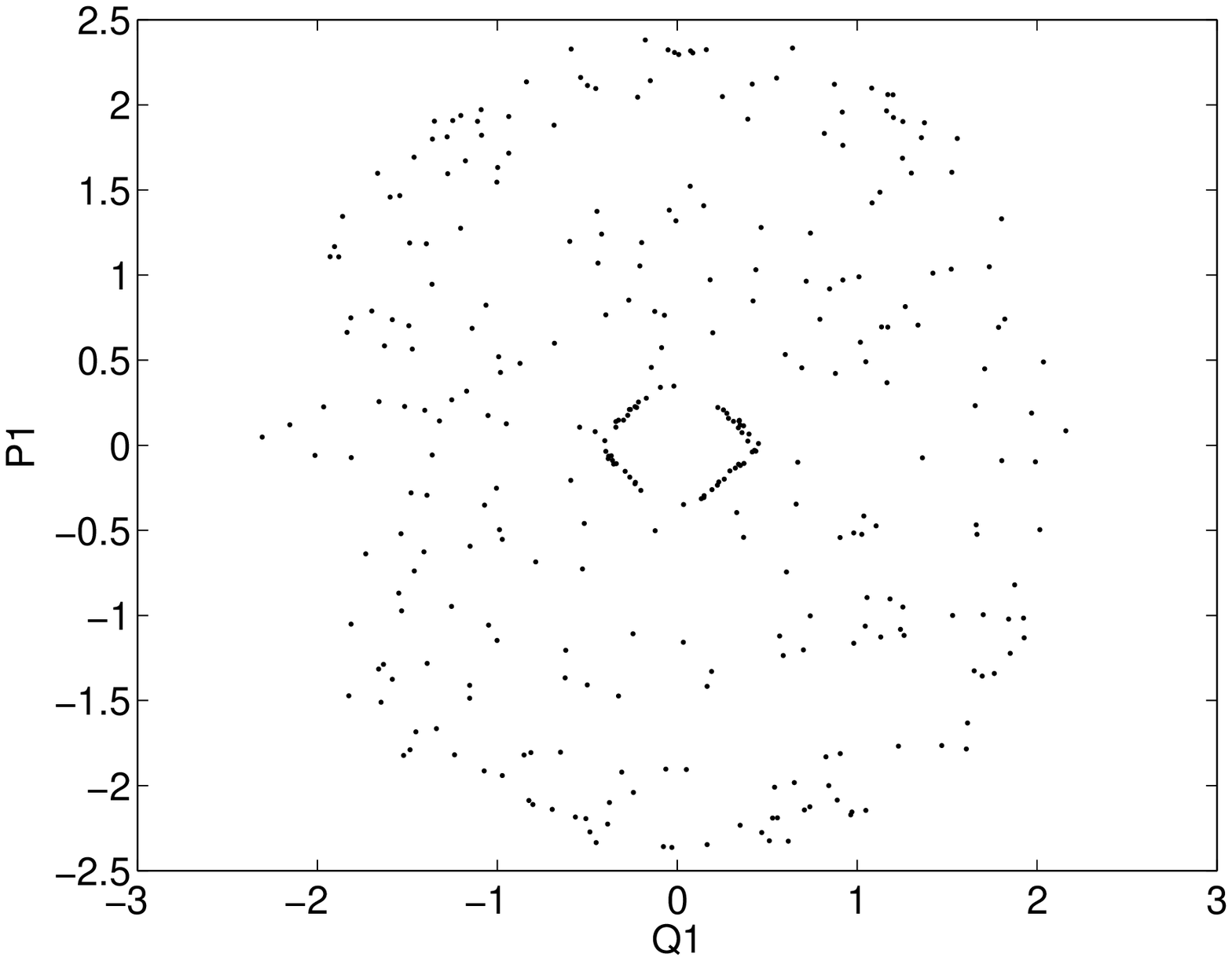}
}
\caption{\label{cap:2DW-model-pot}
Equipotential contours (left) of the model
Hamiltonian ${\cal H}_{0}$ for different energies and the Poincar\'e
section (right) of a selected trajectory at $E=3$. Some tiny quasi-integrable
islands are avoided (mainly at $(0,0)$). }
\end{figure}

In Figure \ref{cap:2DW-model-pot} we display the equipotential contours of the model 
Hamiltonian (\ref{eq:2DW_hamiltonian}) with $x_0=1$. We observe that the equipotential 
surfaces are circles but as the energy is increased they become more and more deformed 
leading to chaotic motion. Our analysis is focused on an energy window around $E \sim 3$ 
where the motion is mainly chaotic. This is illustrated in the right panel of Figure 
\ref{cap:2DW-model-pot} where we report the Poincar\'e section (of the phase space) of 
a selected trajectory, obtained from ${\cal H}_{0}$ at $E=3$. The ergodicity of the motion 
is illustrated by the Poincar\'e section, filling the plane except from some tiny quasi-
integrable islands. 

%%%%  FIGURE %%%%%%%%%%%%%%%%%%%%%%%%%%
%%%% END FIGURE %%%%%%%%%%%%%%%%%%%%%%%

The perturbation is described by ${\cal W} = Q_{1}^{2}Q_{2}^{2}$. 
In the classical analysis there is only one significant regime 
for the strength of the perturbation. Namely, the perturbation is 
considered to be classically small if 
\be{0}
\delta x \ll \varepsilon_{\tbox{cl}}\szk
\ee
where $\varepsilon_{\tbox{cl}}=1$. This is the regime 
where (classical) linear analysis applies. 
Namely, within this regime the deformation of the energy 
surface ${\cal H}_0=E$ can be described as 
a linear process (see Eq.~(\ref{e26})).

%%%%%%%%%%%%%%%%%%%%%%%%%%
\subsection{Energy levels}

Let us now quantize the system. For obvious reasons 
we are considering a de-symmetrized 1/8 well 
with Dirichlet boundary conditions on the
lines $Q_{1}=0$, $Q_{2}=0$ and $Q_{1}=Q_{2}$. 
The matrix representation of ${\cal H}_0$ 
in the basis of the un-coupled system 
is very simple. The eigenstates of the 
Hamiltonian ${\cal H}_0$ are then obtained numerically.

As mentioned above, we consider the experiments 
to take place in an energy window $2.8<E<3.1$ which 
is classically small and where the motion is predominantly 
chaotic. Nevertheless, quantum mechanically, this energy 
window is large, i.e., many levels are found therein. 
The local mean level spacing $\Delta(E)$ at this energy 
range is given approximately by $\Delta\sim4.3\,\hbar^{2}$. 
The smallest $\hbar$ that we can handle is $\hbar=0.012$ 
resulting in a matrix size of about $4000\times4000$. 
Unless stated otherwise, all the numerical data presented 
below correspond to a quantization with $\hbar=0.012$.

As it was previously mentioned in the introduction, the main focus of quantum chaos studies 
has so far been on 
issues of spectral statistics \cite{Sbook98,Hbook00}. In this context it turns out that the sub
-$\hbar$ statistical features of the energy spectrum are "universal", and obey the predictions 
of RMT. In particular we expect that the level spacing distribution $P(s)$ of 
the "unfolded" (with respect to $\Delta$) level spacings $s_{n}=(E_{n+1}-E_{n})/\Delta$ 
will follow with high accuracy the so-called \emph{Wigner} \emph{surmise}. 
For systems with time reversal symmetry it takes the form \cite{Sbook98,BGV84}
\begin{equation}
\label{eq:GOE}
P_{\tbox{spacings}}(s)=\frac{\pi}{2}\, s\,\eexp{-\frac{\pi}{4}\, s^{2}}\szk
\end{equation}
indicating that there is a linear repulsion between nearby levels. 
Non-universal (i.e. system specific) features are reflected only in the 
large scale properties of the spectrum and constitute the fingerprints 
of the underlying classical chaotic dynamics.

\begin{figure}
\centerline{\includegraphics[width=0.6\hsize,clip]{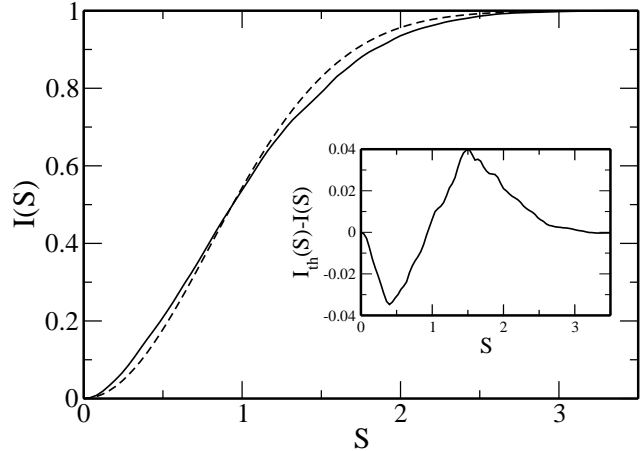}}
\caption{\label{cap:2DW-LSD} 
The integrated level spacing distribution $I(S)$
of the unperturbed Hamiltonian ${\cal H}_{0}$ ($\hbar=0.012$). 
The dashed line is the theoretical prediction for the GOE. 
\emph{Inset}:
Difference between the theoretical prediction $I_{\rm th}(S)$
and the actual distribution $I(S)$.}
\end{figure}

The de-symmetrized 2DW model shows time reversal symmetry, and  
therefore we expect the distribution to follow Eq.(\ref{eq:GOE}). 
The analysis is carried out only for the levels contained 
in the chosen energy window around $E=3$. 
Instead of plotting $P(s)$ we show the integrated 
distribution $I(s)=\int_{0}^{s} P(s')ds'$,  
which is independent of the bin size of the histogram. 
In Figure \ref{cap:2DW-LSD} we present our numerical data 
for $I(s)$ while the inset shows the deviations from 
the theoretical prediction (\ref{eq:GOE}). 
The agreement with the theory is fairly good and the level 
repulsion is clearly observed. The observed 
deviations have to be related on the one hand to the 
tiny quasi-integrable islands that exist at $E=3$ as well 
as to rather limited level statistics.

%%%%  FIGURE %%%%%%%%%%%%%%%%%%%%%%%%%%
%%%% END FIGURE %%%%%%%%%%%%%%%%%%%%%%%

%%%%%%%%%%%%%%%%%%%%%%%%%%%%
\subsection{The band-profile}

In this subsection we explain that the 
band-structure of $\bm{B}$ is related to the 
fluctuations of the classical motion.
This is the major step towards RMT modeling.

Consider a given ergodic trajectory $(Q(t),P(t))$ 
on the energy surface \linebreak ${\cal H}(Q(0),P(0);x_0)=E$.
An example is shown in Fig.~\ref{cap:2DW-model-pot}b. 
We can associate with it a stochastic-like variable  
\begin{equation}
{\cal F}(t) = 
-\frac{\partial{\cal H}}{\partial x}(Q(t),P(t),x(t))\szk
\label{eq:generalized-force}
\end{equation}
which for our linearized Hamiltonian is simply 
the perturbation term \linebreak ${\cal F}=-{\cal W}=-Q_1^2Q_2^2$. 
It can be interpreted as the generalized force 
that acts on the boundary of the 2D well. 
It may have a non-zero average (``conservative" part) 
but below we are interested only in its fluctuations.

\begin{figure}[t]
\centerline{\includegraphics[width=0.6\hsize,clip]{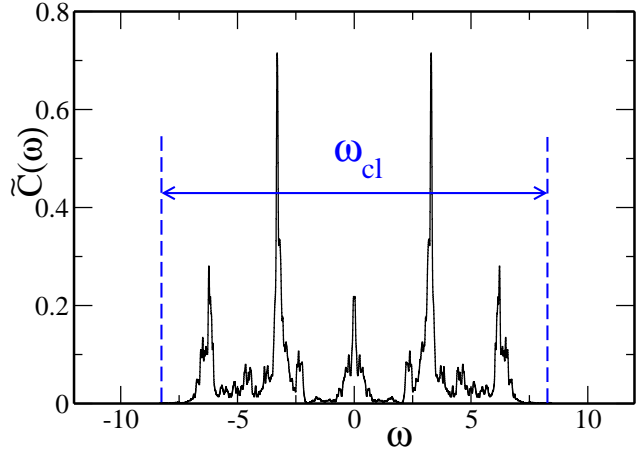}}
\caption{\label{cap:clas-Cw}
The classical power-spectrum of the model (\ref{eq:2DW_hamiltonian}). 
The classical cut-off frequency $\omega_{\tbox{cl}}\simeq 7$ is indicated 
by perpendicular dashed lines. }
\end{figure}

In order to characterize the fluctuations 
of ${\cal F}(t)$ we introduce the autocorrelation 
function $C(\tau)$ 
\begin{equation}
C(\tau)=\langle{\cal F}(t){\cal F}(t+\tau)\rangle
- \langle{\cal F}^2\rangle\szp
\label{eq:autocorrelation}
\end{equation}
The angular brackets denote an averaging which 
is either micro-canonical over some 
initial conditions $\left(Q(0),P(0)\right)$ 
or temporal (due to the assumed ergodicity). 
The power spectrum for the 2D well model
is shown in Fig.\ref{cap:clas-Cw} (see solid line).

For generic chaotic systems (described by smooth Hamiltonians), 
the fluctuations are characterized by a short 
correlation time $\tau_{\tbox{cl}}$,  
after which the correlations are negligible. 
In generic circumstances $\tau_{\tbox{cl}}$ 
is essentially the ergodic time. 
For our model system $\tau_{\tbox{cl}}\sim 1$.

The power spectrum of the fluctuations $\tilde{C}(\omega)$ 
is defined by a Fourier transform:
\begin{equation}
\label{powsp}
\tilde{C}(\omega) =
\int_{-\infty}^{\infty} C(\tau) \exp(i\omega\tau)d\tau\szp
\end{equation}
This power spectrum is characterized 
by a cut-off frequency $\omega_{\tbox{cl}}$ which is 
inverse proportional to the classical correlation time
\begin{equation} 
\omega_{\tbox{cl}}=\frac{2\pi}{\tau_{\tbox{cl}}} \szp
\label{eq:omega_cl}
\end{equation}
Indeed in the case of our model system we get  
$\omega_{\tbox{cl}}\sim7$ which is in agreement 
with Fig.\ref{cap:clas-Cw}.

%%%%  FIGURE %%%%%%%%%%%%%%%%%%%%%%%%%%
%%%% END FIGURE %%%%%%%%%%%%%%%%%%%%%%%

The implication of having a short 
but non-vanishing classical correlation 
time $\tau_{\tbox{cl}}$  is having large 
but finite bandwidth in the perturbation 
matrix $\bm{B}$. This follows from the identity 
\be{0}
\tilde{C}(\omega) = 
\sum_n |B_{nm}|^2 2\pi \delta\left(\omega-\frac{E_n-E_m}{\hbar}\right)\szk
\ee
which implies
\begin{equation}
\label{eq:bandprofile}
\langle|\bm{B}_{nm}|^{2}\rangle=\frac{\Delta} {2\pi\hbar}
\tilde{C}\left(\omega=\frac{E_n-E_m}{\hbar}\right)\szp
\end{equation}
Hence the matrix elements of the perturbation matrix $\bm{B}$ 
are extremely small outside of a band 
of width $b=\hbar\omega_{\tbox{cl}}/\Delta$.

In the inset of Figure \ref{cap:band-profile} 
we show a snapshot of the perturbation matrix $|\bm{B}_{nm}|^2$. 
It clearly shows a band-structure. At the same figure we 
also display the band-profiles for different values of $\hbar$. 
A good agreement with the classical power 
spectrum ${\tilde C}(\omega)$ is evident.

It is important to realize that upon quantization we end up 
with {\em two} distinct energy scales. One is obviously 
the mean level spacing (see previous subsection) 
\be{0}
\label{mlsn1}
\Delta \propto \hbar^d \szk
\ee
where the dimensionality is $d=2$ in case of our 
model system. The other energy scale is the bandwidth 
\be{0}
\label{blsn1}
\Delta_b = \frac{2\pi\hbar}{\tau_{\tbox{cl}}} = b\Delta \szp
\ee
This energy scale is also known in the corresponding literature 
as the "non-universal" energy scale \cite{Bbook91}, 
or (in case of diffusive motion) as the Thouless energy \cite{Ibook97}
{\footnote{The dimensionless parameter $b$ scales like 
$b\propto \hbar^{-(d-1)}$ and in the frame of mesoscopic systems 
is recognized as the dimensionless Thouless conductance\cite{Ibook97}.}. 
One has to notice that deep in the semiclassical limit $\hbar\rightarrow0$
these two energy scales differ enormously from one another
(provided $d\geq 2$). We shall see in the following sections that 
this scale separation has dramatic consequences 
in the theory of driven quantum systems.

%%%%  FIGURE %%%%%%%%%%%%%%%%%%%%%%%%%%
\begin{figure}[t]
\centerline{\includegraphics[width=0.70\hsize,clip]{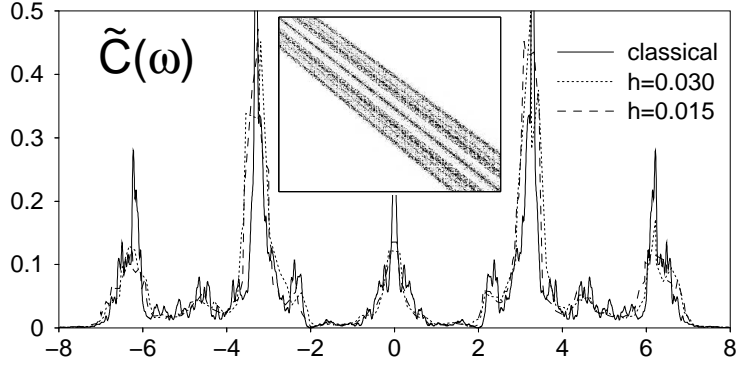}}
\caption{\label{cap:band-profile}
The band-profile $(2\pi\hbar/\Delta)\cdot|\bm{B}_{nm}|^{2}$
versus $\omega=(E_{n}-E_{m})/\hbar$ is compared with the classical power spectrum $\tilde{C}
(\omega)$. Inset: a snapshot of the perturbation matrix $\bm{B}$.}
\end{figure}
%%%% END FIGURE %%%%%%%%%%%%%%%%%%%%%%%

%%%%%%%%%%%%%%%%%%%%%%%%%%%%%%%%%%%%%%%%
\subsection{Distribution of couplings}

We investigate further the statistical 
properties of the matrix elements $\bm{B}_{nm}$ of the perturbation matrix, by studying 
their distribution. RMT assumes that upon appropriate unfolding they must be distributed in a 
Gaussian manner. The 'unfolding' aims to remove system specific properties and reveal the 
underlying universality. It is done by normalizing the matrix elements with the local standard 
deviation $\sigma =\sqrt{\langle|{\bf B}_{nm}|^2\rangle}$ related through Eq.~(\ref{eq:bandprofile}) 
with the classical power spectrum ${\tilde C}(\omega)$.

The existing literature is not conclusive about the distribution of the normalized matrix elements 
$q=\bm{B}_{nm}/\sigma$. Specifically, Berry \cite{B77} and more recently Prosen \cite{PR93,P94}, 
claimed that ${\cal P}(q)$ should be Gaussian. On the other hand, Austin and Wilkinson \cite{AW92}
have found that the Gaussian is approached only in the limit of high quantum numbers while for
small numbers, i.e., low energies, a different distribution applies,
namely 
\begin{equation}
P_{\tbox{couplings}}(q)
=\frac{\Gamma(\frac{N}{2})}{\sqrt{\pi N}\Gamma(\frac{N-1}{2})}
\left(1-\frac{q^{2}}{N}\right)^{(N-3)/2}\szp
\label{eq:wilkinson}
\end{equation}
This is the distribution of the elements of an $N$-dimensional vector, distributed randomly over 
the surface of an $N$-dimensional sphere of radius $\sqrt{N}$. For $N\rightarrow\infty$ 
this distribution approaches a Gaussian. 

The distribution ${\cal P}(q)$ for our model is reported in Figure \ref{cap:2DWmat_dist}. The solid 
line corresponds to a Gaussian of unit variance while the dashed-dotted line is obtained by fitting 
Eq.~(\ref{eq:wilkinson}) to the numerical data using $N$ as a fitting parameter. We observe that the 
Gaussian resembles better our numerical data although deviations, especially for matrix elements 
close to zero, can be clearly seen. 
We attribute these deviations to the existence of the tiny stability islands in the phase space. 
Trajectories started in those islands cannot reach the chaotic sea and vice versa. Quantum mechanically 
the consequence of this would be vanishing matrix elements $\bm{B}_{nm}$ which represent the classically 
forbidden transitions.

%%%%  FIGURE %%%%%%%%%%%%%%%%%%%%%%%%%%
\begin{figure}
\centerline{\includegraphics[width=0.6\hsize,clip]{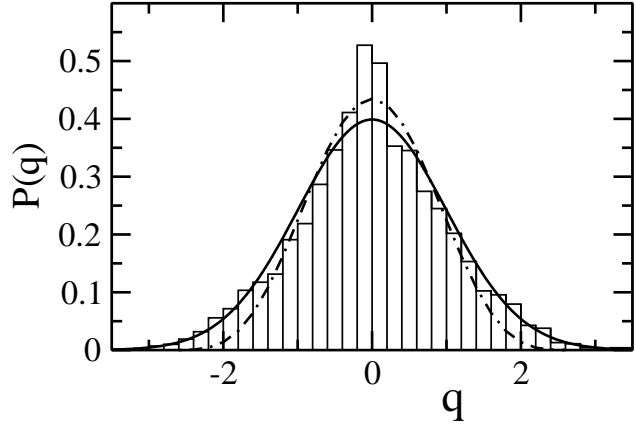}}
\caption{\label{cap:2DWmat_dist}
Distribution of matrix elements $q$ around
$E=3$ rescaled with the averaged band-profile . The solid black line
corresponds to a Gaussian distribution with unit variance while the
dashed-dotted line corresponds to a fit from Eq.~(\ref{eq:wilkinson}) with
a fitting parameter $N=7.8$. The quantization corresponds to $\hbar=0.03$.}
\end{figure}
%%%% END FIGURE %%%%%%%%%%%%%%%%%%%%%%%

%%%%%%%%%%%%%%%%%%%%%%%%%%%%%%%%%%%%%%%%%%%%%%%%%%%%%%%%%%
\subsection{RMT modeling \label{subsec:RMT} }

It was the idea of Wigner \cite{W55,W57} more than forty years ago, to study a simplified 
model, where the Hamiltonian is given by Eq.~(\ref{eq:2DW_hamilton_matrx}), and where 
$\bm{B}$ is a Banded {\it Random} Matrix (BRM) \cite{FLW91,FGIM93,FCIC96}. The diagonal 
matrix $\bm{E}$ 
has elements which are the ordered energies $\{E_n\}$, with mean level spacing $\Delta$. 
The perturbation matrix $\bm{B}$ has a \emph{rectangular} band-profile of band-size $b$. 
Within the band $0<|n-m| \leq b$ the elements are independent random variables given by
a Gaussian distribution with zero mean and a variance $\sigma^2=\langle|\bm{B}_{nm}|^{2}
\rangle$. Outside the band they vanish. We refer to this model as the \emph{Wigner BRM} 
model (WBRM).

Given the band-profile, we can use Eq.(\ref{eq:bandprofile}) in reverse direction to 
calculate the correlation function $C(\tau)$. For the WBRM model we get
\begin{equation}
\label{wbrmct}
C(\tau)=2\sigma^{2}b\,{\rm sinc}\left(\tau/\tau_{\tbox{cl}}\right)\szk
\end{equation}
where $\tau_{\tbox{cl}}= \hbar/\Delta_b$. Thus, there are three parameters $(\Delta,b,\sigma)$ 
that define the WBRM model.

The WBRM model can be regarded as a {\em simplified} local description of a true Hamiltonian 
matrix. This approach is attractive both analytically and numerically. Analytical calculations 
are greatly simplified by the assumption that the off-diagonal terms can be treated as 
independent random numbers. Also from a numerical point of view it is quite a tough task to 
calculate the true matrix elements of the $\bm{B}$ matrix. It requires a preliminary step 
where the chaotic ${\cal H}_0$ is diagonalized. Due to memory limitations one ends up with 
quite small matrices. For the Pullen-Edmonds model we were able to handle matrices of final 
size $N=4000$ maximum. This should be contrasted with the WBRM simulations, where using self
-expanding algorithm \cite{IKPT97,CIK00} we were able to handle system sizes up to $N=100000$ along 
with significantly reduced CPU time.

We would like to stress again that the underlying assumption of WBRM, 
namely that the off-diagonal elements are {\em uncorrelated} random numbers, 
has to be treated with extreme care. 

The WBRM model involves an additional simplification. Namely, one assumes 
that $\bm{B}$ has a \emph{rectangular} band-profile.  
A simple inspection of the band-profile of our model Eq.~(\ref{eq:2DW_hamiltonian}) 
shows that this is not the case (see Fig.~\ref{cap:band-profile}). 
We eliminate this simplification by introducing
a RMT model that is even closer to the dynamical one. 
To this end, we randomize the signs of the off-diagonal elements 
of the perturbation matrix $\bm{B}$ keeping its band-structure 
intact. This procedure leads to a random model that exhibits 
only universal properties while it lack any semiclassical limit. 
We will refer to it as the \emph{effective} banded random matrix model (EBRM).

%%%%%%%%%%%%%%%%%%%%%%%%%%%%%%%%%%%%%%%%%%%%%%%%%%%%%%%%%%%%%%%%%%%%
\section{The Parametric Evolution of the Eigenfunctions \label{sec:regimes}}

As we change the parameter $\delta x$ in the Hamiltonian Eq.~(\ref{eq:2DW_hamilton_matrx}), 
the instantaneous eigenstates $\{|n(x)\rangle\}$  evolve and undergo 
structural changes. In order to understand the actual dynamics, it is important to understand 
these structural changes. This leads to the introduction of 
\begin{equation}
\label{ef}
P(n|m) = |\langle n(x)|m(x_0)\rangle|^2 \szk
\end{equation}
which is easier to analyze than $P_t(n|n_0)$.
Up to some trivial scaling and shifting $P(n|m)$ is essentially the local density of states (LDoS):
\be{0}
P(E|m)=\sum_n|\langle n(x)|m(x_0) \rangle|^2\delta(E-E_n)\szp
\ee
The averaged distribution $P(r)$ is defined in  
complete analogy with the definition of $P_t(r)$.  
Namely, we use the notation ${r=n-m}$, and average 
over several $m$ states with 
roughly the same energy $E_m\sim E$.

Generically $P(r)$ undergoes 
the following structural changes 
as a function of growing $\delta x$. 
We first summarize the generic picture, 
which involves the parametric scales 
$\varepsilon_c$ and $\varepsilon_{\tbox{prt}}$.
and the approximations 
$P_{\tbox{FOPT}}$, $P_{\tbox{prt}}$, and  $P_{\tbox{sc}}$. 
Then we discuss how to determine these scales,  
and what these approximations are.

\begin{itemize}

\item 
The first order perturbative theory (FOPT) regime 
is defined as the range $\delta x < \varepsilon_c$ 
where we can use FOPT to get 
an approximation that we denote 
as $P() \approx P_{\tbox{FOPT}}$. 

\item
The (extended) perturbative regime is defined 
as the range   $\varepsilon_c <\delta x < \varepsilon_{\tbox{prt}}$
where we can use perturbation theory (to an infinite order) 
to get a meaningful approximation  
that we denote as $P()\approx P_{\tbox{prt}}$. 
Obviously $P_{\tbox{prt}}$ reduces to $P_{\tbox{FOPT}}$ 
in the FOPT regime.

\item
The non-perturbative regime 
is defined as the range  
$\delta x > \varepsilon_{\tbox{prt}}$ 
where perturbation theory becomes 
non-applicable. In this regime 
we have to use either RMT 
or semiclassics in order to get 
an approximation that we 
denote as $P()\approx P_{\tbox{sc}}$. 

\end{itemize}

Irrespective of these structural changes, it can be 
proved that the variance of $P(r)$ is strictly linear 
and given by the expression 
\be{26}
\delta E(\delta x) \ = 
\ \sqrt{C(0)} \ \delta x 
\ \equiv \  \delta E_{\tbox{cl}}\szp
\ee
The only assumption that underlines this 
statement is $\delta x \ll \varepsilon_{\tbox{cl}}$. 
It reflects the linear departure 
of the energy surfaces.

%%%%%%%%%%%%%%%%%%%%%%%%%%%%%%%%%5
\subsection{Approximations for $P(n|m)$}

The simplest regime is obviously the FOPT regime where, for $P(n|m)$, we can use the standard textbook 
approximation $P_{\tbox{FOPT}}(n|m)\approx 1$ for $n=m$, 
while  
\begin{equation}
\label{Pst1}
P_{\tbox{FOPT}}(n|m) = \frac{\delta x^2 \ |\bm{B}_{nm}|^2}{(E_n{-}E_m)^2}\szk
\end{equation}
for $n\ne m$. If outside of the band 
we have $\bm{B}_{nm} = 0$, as in the WBRM model, 
then $P_{\tbox{FOPT}}(r)=0$ for $|r|>b$.
To find the higher order tails 
(outside of the band) we have to go to higher 
orders in perturbation theory.
Obviously this approximation makes sense only 
as long as $\delta x < \varepsilon_c$ where 
\begin{equation}
\label{ecpn1}
\varepsilon_c = \Delta / \sigma \sim \hbar^{(1+d)/2}\szk
\end{equation}
and $d$ is the degrees of freedom of our system ($d=2$ for the 2D well model).

If $\delta x > \varepsilon_c$ 
but not too large then we still have tail regions 
which are described by FOPT. 
This is a non-trivial observation 
which can be justified by using perturbation 
theory to infinite order. Then we can argue that 
a reasonable approximation is 
\begin{equation}
\label{Plor}
P_{\tbox{prt}}(n|m) = \frac{\delta x^2 \ |\bm{B}_{nm}|^2}{(E_n{-}E_m)^2 + \Gamma^2}\szk
\end{equation}
where $\Gamma$ is evaluated by imposing normalization of $P_{\rm prt}(n|m)$. In the case of
WBRM model $\Gamma=  (\sigma \delta x/\Delta)^2\times\Delta$. The appearance of $\Gamma$ in 
the above expression cannot be obtained from any {\it finite-order} perturbation theory: 
Formally it requires summation to infinite order. Outside of the bandwidth the tails decay 
faster than exponentially. Note that $P_{\tbox{prt}}(n|m)$ is 
a Lorentzian in the case of a flat bandwidth (WBRM model), 
while in the general case it can be described as a "core-tail" structure.

Obviously the above approximation makes sense only 
as long as $\Gamma(\delta x) < \Delta_b$. 
This expression assumes that the bandwidth $\Delta_b$ 
is sharply defined, as in the WBRM model. 
By elimination this leads to the determination 
of $\varepsilon_{\tbox{prt}}$, which in case of the WBRM model is simply
\begin{equation}
\label{eprtn1}
\varepsilon_{\tbox{prt}} \,\,=\,\, \sqrt{b} \,\, 
\varepsilon_c\sim {\hbar\over \tau_{\tbox{cl}}{\sqrt{C(0)}}}\szp
\end{equation}
In more general cases the bandwidth is not sharply defined. 
Then we have to define the perturbative regime using 
a practical numerical procedure. The natural definition that 
we adopt is as follows. 
We calculate the spreading $\delta E(\delta x)$, 
which is a linear function. 
Then we calculate $\delta E_{\tbox{prt}}(\delta x)$,   
using Eq.(\ref{Plor})). This quantity always 
saturates for large $\delta x$ 
because of having finite bandwidth.
We compare it to the exact $\delta E(\delta x)$, 
and define $\varepsilon_{\tbox{prt}}$ for instance as the $80\%$ departure point.

%%%%  FIGURE %%%%%%%%%%%%%%%%%%%%%%%%%%
\begin{figure}[p]
\centerline{\includegraphics[height=0.85\vsize,clip]{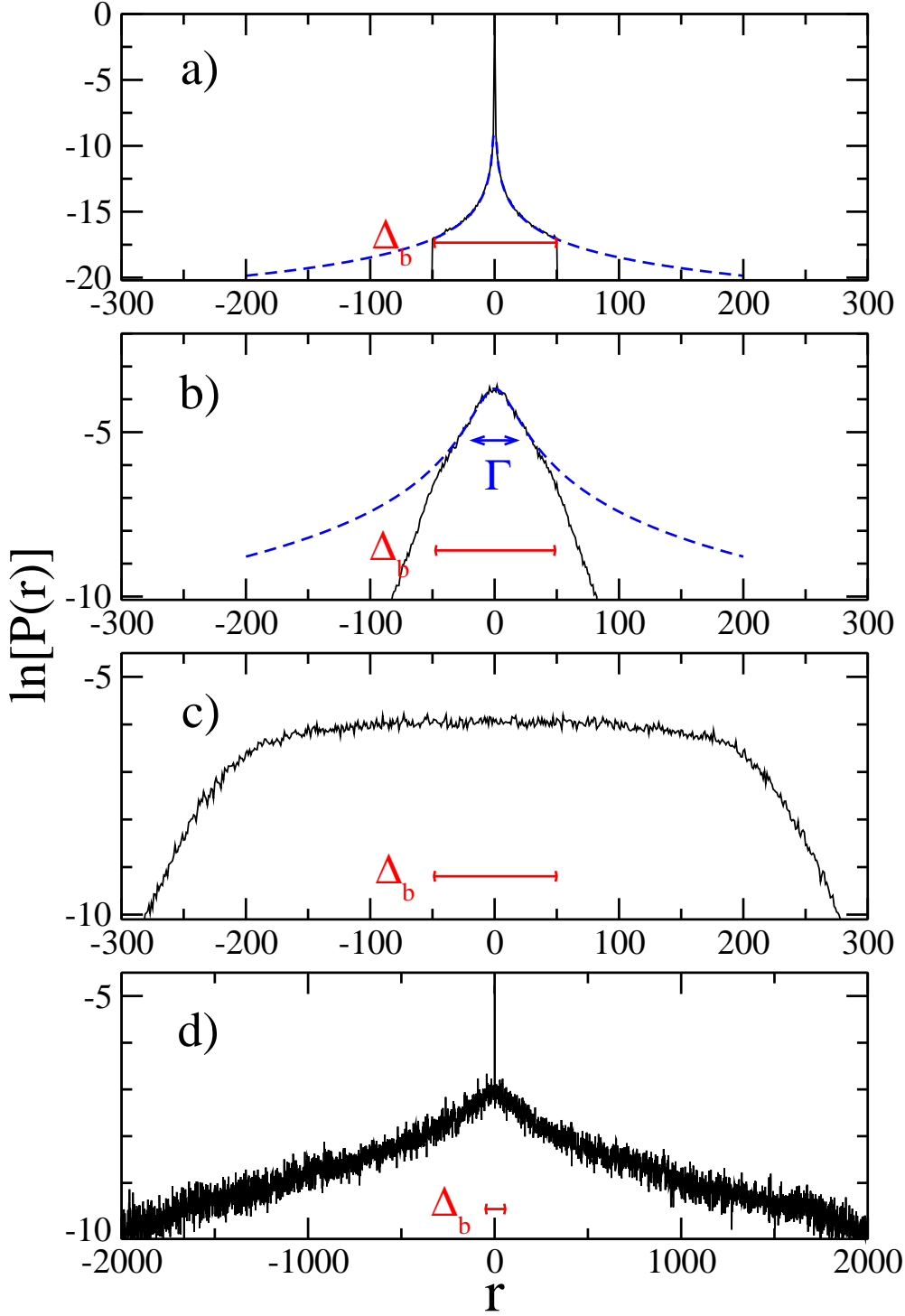}}
\caption{\label{cap:WBRM}
The parametric evolution of eigenstates of a WBRM model with $\sigma=1$ and $b=50$: 
(a) Standard perturbative regime corresponding to $\epsilon=0.01$, (b) Extended 
perturbative regime with $\epsilon=2$ (c) Non-perturbative (ergodic) regime with 
$\epsilon= 12$ and (d) localized regime with $\epsilon=1$. In (a-c) the mean level spacing
$\Delta=1$ while in (d) $\Delta=10^{-3}$. The bandwidth $\Delta_b=\Delta \times b$
is indicated in all cases. In (b) the blue dashed line corresponds to a Lorentzian with
$\Gamma\approx 16\ll \Delta_b$ while in (a) we have $\Gamma\approx 10^{-4}\ll \Delta$
which therefore reduces to the standard FOPT result.}
\end{figure}
%%%% END FIGURE %%%%%%%%%%%%%%%%%%%%%%%

%%%%  FIGURE %%%%%%%%%%%%%%%%%%%%%%%%%%
\begin{figure}[t]
\centerline{\includegraphics[height=0.6\vsize,angle=0,clip]{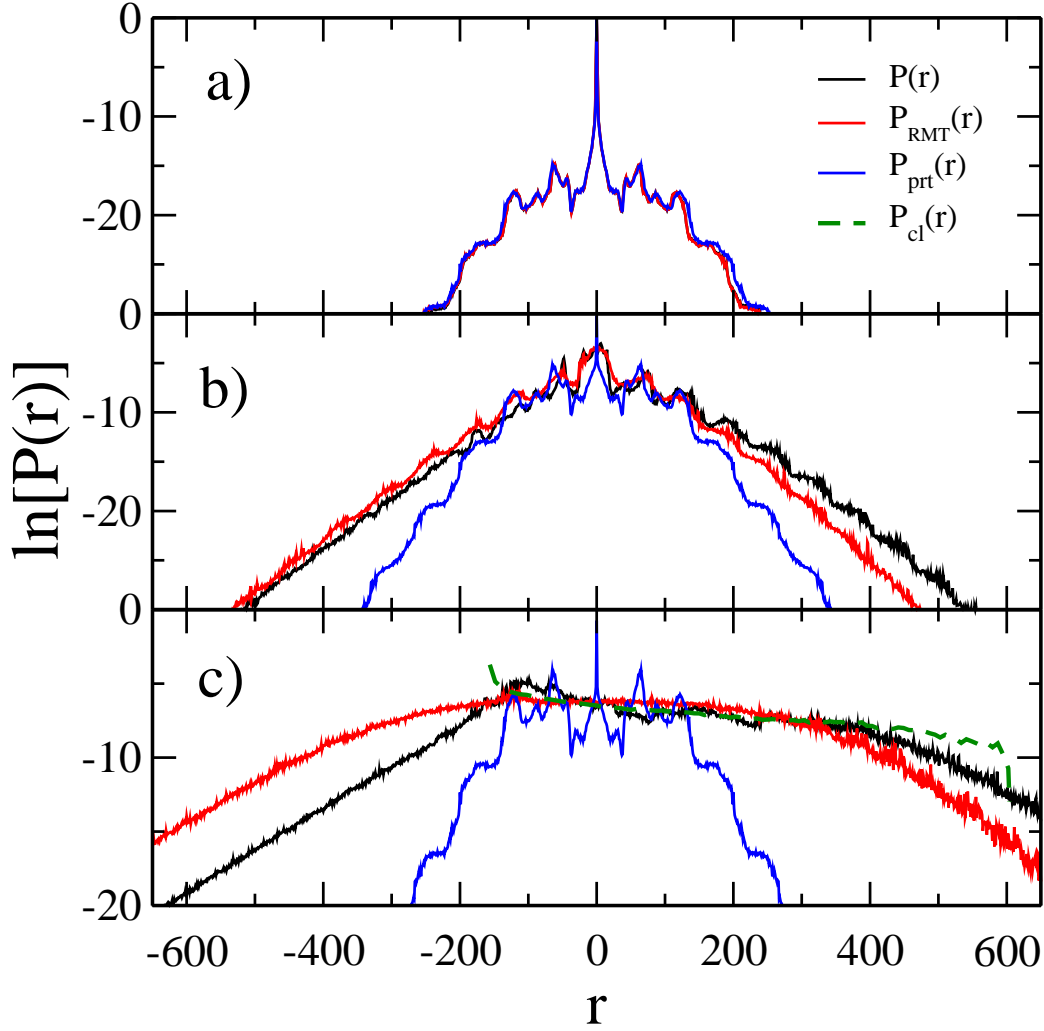}}
\caption{\label{cap:LDOS2dw}
The quantal profile $P(r)$ for the 2DW model is compared with
$P_{\tbox{prt}}(r)$ and with the corresponding $P_{\tbox{RMT}}(r)$
of the EBRM model. The perturbation strength $\delta x$ is in (a)
$\delta x=0.00035$, (b) $\delta x= 0.04945$ and (c) $\delta x = 0.29$.
We are using here the $\hbar=0.012$ output. In the lower plot the 
classical LDoS profile $P_{\tbox{cl}}(r)$
is represented by a green heavy dashed line. } 
\end{figure}
%%%% END FIGURE %%%%%%%%%%%%%%%%%%%%%%%

What happens if perturbation theory completely fails? 
In the WBRM model the LDoS becomes semicircle:

\begin{equation}
\label{Psc}
P_{\tbox{sc}}(n|m) \ \ = \ \
\frac{1}{2\pi\Delta} \sqrt{4-\left( \frac{E_n-E_m}{\Delta}\right)^2}\szk
\end{equation}
while in systems that have a semiclassical limit 
we expect to get 
\begin{equation}
\label{Pcl}
P_{\tbox{sc}}(n|m) \ = \ \int {dQdP\over (2 \pi\hbar)^d}  \rho_n(Q,P)\rho_m(Q,P)\szk
\end{equation}
where $\rho_m(Q,P)$ and $\rho_n(Q,P)$ 
are the Wigner functions that correspond to 
the eigenstates $|m(x_0)\rangle$ and $|n(x)\rangle$ respectively.

%%%%%%%%%%%%%%%%%%%%%%%%%%%%%%%%%
\subsection{The $P(n|m)$ in practice}

There are some findings that go beyond the above generic picture and, for completeness, 
we mention them. The first one is the "localization regime" which is found in the case 
of the WBRM model for  $\varepsilon > \varepsilon_{\tbox{loc}}$.  where
\be{0}
\varepsilon_{\tbox{loc}} = b^{\rm 3/2} \varepsilon_c\szp
\ee
In this regime it is important 
to distinguish between the non-averaged $P(n|m)$ and the averaged $P(r)$ because the eigenfunctions 
are non-ergodic but rather localized. This localization is not reflected in the LDoS which is 
still a semicircle. A typical eigenstate is exponentially localized within an energy range $\delta 
E_{\xi} = \xi\Delta$ much smaller than $\delta E_{\tbox{cl}}$. 
The localization length is $\xi \approx b^2$. 
In actual physical applications it is not clear whether there is such a type of localization. 
The above scenario for the WBRM model is summarized in Fig.~\ref{cap:WBRM} where we plot $P(n|m)$ 
in the various regimes. The localized regime is not an issue in the present work and therefore we 
will no further be concerned with it.

The other deviation from the generic scenario, is the appearance 
of a non-universal "twilight regime" which can be found 
for some quantized systems \cite{MKC04}. In this regime a co-existence 
of a perturbative and a semiclassical structure can be observed. 
For the Pullen-Edmonds model (\ref{eq:2DW_hamiltonian}) there is no such distinct regime.

For the Hamiltonian model described by Eq.~(\ref{eq:2DW_hamiltonian}) the borders between 
the regimes can be estimated \cite{CK01}. Namely  $\varepsilon_c\approx 3.8 \hbar^{3/2}$  
and $\varepsilon_{\tbox{prt}}\approx 5.3 \hbar$.
In Fig.~\ref{cap:LDOS2dw} we report the parametric evolution of the eigenstates for the Hamiltonian 
model of Eqs.~(\ref{eq:2DW_hamiltonian}}) and we compare the outcomes with the results of the EBRM 
model \cite{CK01}. Despite the overall quantitative agreement, some differences can be detected:
\begin{itemize}
\item In the FOPT regime (see Fig.~\ref{cap:LDOS2dw}a), the RMT strategy fails in the 
far tails regime $\Delta \times |r|> \Delta_b$ where system specific interference phenomena 
become important.

\item In the extended perturbative regime (see Fig.~\ref{cap:LDOS2dw}b) the line-shape of the 
averaged wavefunction $P(n|m)$ is different from Lorentzian. Still the general features of 
$P_{\tbox{prt}}$ (core-tail structure) can be detected. In a sense, Wigner's Lorentzian (\ref{Plor}) 
is a special case of core-tail structure. Finally, as in the standard perturbative regime one 
observes that the far-tails are dominated by either destructive interference (left tail), or 
by constructive interference (right tail).

\item Deep in the non-perturbative regime ( $\varepsilon>\varepsilon_{\tbox{prt}}$ ) 
the overlaps $P(n|m)$ are well approximated by the semiclassical expression. 
The exact shape is determined by simple classical considerations \cite{CK01,BGI98}. 
This is in contrast to the WBRM model which does not have a classical limit. 
\end{itemize}

%%%%%%%%%%%%%%%%%%%%%%%%%%%%%%%%%%%%%%%%%%%%%%%%%%%%%%%%%%%%%%%%%%%
\section{Linear Response Theory \label{sec:Meas_of_QI}}

The definition of regimes for driven systems is more complicated than the corresponding 
definition in case of LDoS theory. It is clear that for short times we always can use 
time-dependent FOPT. The question is, of course, what happens next. There we have to 
distinguish between two types of scenarios. One type of scenario 
is {\em wavepacket dynamics} for which the dynamics is a transient from  
a preparation state to some new ergodic state. 
The second type of scenario is {\em persistent driving}, either linear driving 
($\dot{x} = \varepsilon$) or periodic driving ($x(t)=\varepsilon\sin(\Omega t)$). 
In the latter case the strength of the perturbation depends 
on the rate of the driving, not just on the amplitude. 
The relevant question is {\em whether the long time dynamics can be deduced from the 
short time analysis}. To say that the dynamics is of perturbative nature means that 
the short time dynamics can be deduced from FOPT, while the long time dynamics can be 
deduced on the basis of a Markovian (stochastic) assumption. The best known example is 
the derivation of the exponential Wigner law for the decay of metastable state. 
The Fermi-Golden-Rule (FGR) is used to determine the initial rate for the escaping process, and then 
the long-time result is extrapolated by assuming that the decay proceeds in 
a stochastic-like manner. Similar reasoning is used in deriving 
the Pauli master equation which is used to describe the stochastic-like transitions 
between the energy levels in atomic systems.

A related question to the issue of regimes is the validity of Linear Response Theory (LRT). 
In order to avoid ambiguities we adopt here a practical definition. Whenever the result 
of the calculation depends only on the two point correlation function $C(\tau)$, 
or equivalently only on the band-profile of the perturbation 
(which is described by $\tilde{C}(\omega)$), then we refer to it as "LRT". 
This implies that higher order correlations are not expressed. 
There is a (wrong) tendency to associate LRT with FOPT. In fact the validity 
of LRT is not simply related to FOPT. We shall clarify this issue in the next section.

For both $\delta E(t)$ and ${\cal P}(t)$ we have "LRT formulas" which we discuss 
in the next sections. Writing the driving pulse as $\delta x(t) = \varepsilon f(t)$ 
for the spreading we get: 
\begin{equation} 
\label{e1}
\delta E^2(t) \ = 
\ \varepsilon^2 \times
\int_{-\infty}^{\infty} 
\frac{d\omega}{2\pi}
\tilde{C}(\omega) 
\tilde{F}_t(\omega)\szk
\end{equation}
while for the survival probability we have
\begin{equation}
{\cal P}(t) = \exp\left( 
-\varepsilon^2 \times
\int_{-\infty}^{\infty}
\frac{d{\omega}}{2\pi}
\tilde{C}(\omega)
\frac{\tilde{F}_{t}(\omega)}{(\hbar\omega)^{2}}
\right)\szp
\label{eq:LRT-qm-P}
\end{equation}
Two spectral functions are involved: 
One is the power spectrum $\tilde{C}(\omega)$ 
of the fluctuations defined in Eq.~(\ref{powsp}), 
and the other $\tilde{F}_t(\omega)$ 
is the spectral content of the driving pulse 
which is defined as 
\begin{equation}
\tilde{F}_{t}(\omega)
=\left|\int_{0}^{t}d{t'}\dot{f}(t')
{\rm e}^{-i\omega t'}\right|^{2}\szp
\label{eq:spectral-content-F}
\end{equation}

Here we summarize the main observations regarding the 
nature of wavepacket dynamics in the various regimes: 

\begin{itemize}

\item 
{\it FOPT regime}: 
In this regime ${\cal P}(t)\sim 1$ for all time, 
indicating that all probability is all the time 
concentrated on the initial level. 
An alternative way to identify this regime is
from $\delta E_{\tbox{core}}(t)$ 
which is trivially equal to $\Delta$.

\item 
{\it Extended perturbative regime}: 
The appearance of a core-tail structure which 
is characterized by separation of scales 
$\Delta\ll \delta E_{\tbox{core}}(t) \ll \delta E(t) \ll \Delta_{b}$. 
The core is of non-perturbative nature,  
but the variance $\delta E^2(t)$ 
is still dominated by the tails. 
The latter are described by perturbation theory.  

\item 
{\it Non-perturbative regime}: 
The existence of this regime is associated 
with having the finite energy scale~$\Delta_b$. 
It is characterized by 
$\Delta_{b} \ll \delta E_{\tbox{core}}(t) \sim \delta E(t)$. 
As implied by the terminology, 
perturbation theory (to any order) 
is not a valid tool for the analysis of 
the energy spreading. Note that in this regime, 
the spreading profile is characterized 
by a single energy scale ($\delta E \sim \delta E_{\tbox{core}}$).

\end{itemize}

%%%%%%%%%%%%%%%%%%%%%%%%%%%%%%%%%%%%%%%%%%%%%%%%%%
%%%%%%%%%%%%%%%%%%%%%%%%%%%%%%%%%%%%%%%%%%%%%%%%%%
\subsection{The energy spreading $\delta E(t)$ \label{sec:LRT}}

Of special importance for understanding quantum dissipation is the theory for the variance 
$\delta E^2(t)$ of the energy spreading. Having  $\delta E(t) \propto \epsilon$
means {\em linear response}. If $\delta E(t)/\epsilon$ depends on $\epsilon$,
we call it ``non-linear response". In this paragraph we explain
that linear response theory (LRT) is based on the ``LRT formula" Eq.(\ref{e1}) 
for the spreading. This formula has a simple classical derivation (see Subsection \ref{sub:LRT-cl} below).

From now on it goes without saying that we assume the {\em classical} conditions
for the validity of Eq.(\ref{e1}) are satisfied (no $\hbar$ involved in such conditions).
The question is {\em what happens to the validity of LRT once we ``quantize" the system}.
In previous publications\cite{CK00,KC03,CK03,CK01}, we were able to argue the following: \\
\hspace*{3mm}\begin{minipage}{0.98\hsize}
\vspace*{0.2cm}
\begin{itemize}
\setlength{\itemsep}{0cm}
\item[(A)]
The LRT formula can be trusted
in the perturbative regime, with the exclusion
of the adiabatic regime.
\item[(B)]
In the sudden limit the LRT formula can
also be trusted in the non-perturbative regime.
\item[(C)]
In general the LRT formula cannot be
trusted in the non-perturbative regime.
\item[(D)]
The LRT formula can be trusted deep in the non-perturbative
regime, provided the system has a classical limit.
\end{itemize}
\vspace*{0.0cm}
\end{minipage}

For a system that does not have a classical limit (Wigner model) we were able to demonstrate 
\cite{CK00,KC03,CK03} that LRT fails in the non-perturbative regime. Namely, for the WBRM model the response 
$\delta E(t)/\epsilon$ becomes $\epsilon$ dependent for large $\epsilon$, meaning that the 
response is non-linear. Hence the statement in item (C) above has been established. We had 
argued that the observed non-linear response is the result of a quantal non-perturbative effect.
Do we have a similar type of non-linear response in the case of {\em quantized chaotic} systems? 
The statement in item (D) above seems to suggest that the observation of such non-linearity 
is not likely. Still, it was argued in \cite{CK03} that this does not exclude the possibility 
of observing a ``weak" non-linearity.

The immediate (naive) tendency is to regard LRT as the outcome of quantum mechanical first 
order perturbation theory (FOPT). In fact the regimes of validity of FOPT and of LRT do not 
coincide. On the one hand we have the adiabatic regime where FOPT is valid as a leading order
description, but not for response calculation. On the other hand, the validity of Eq.(\ref{e1})
goes well beyond FOPT. This leads to the (correct) identification \cite{C00,CK00,CK03} of what we 
call the ``perturbative regime". The border of this regime is determined by the energy scale 
$\Delta_b$, while $\Delta$ is not involved. Outside of the perturbative regime we cannot trust 
the LRT formula. However, as we further explain below, the fact that Eq.(\ref{e1}) is not valid 
in the non-perturbative regime, does not imply that it {\em fails} there.

We stress again that one should  distinguish between ``non-perturbative response" and ``non-linear 
response". These are not  synonyms. As we explain in the next paragraph, the adiabatic regime is 
``perturbative" but ``non-linear", while the semiclassical limit is ``non-perturbative" but ``linear".

In the {\em adiabatic regime}, FOPT implies zero probability to make a transitions to other levels. 
Therefore, to the extent that we can trust the adiabatic approximation, all probability remains 
concentrated on the initial level. Thus, in the adiabatic regime, Eq.(\ref{e1}) is not a valid 
formula: It is essential to use higher orders of perturbation theory, and possibly non-perturbative
corrections (Landau-Zener \cite{W87,W88}), in order to calculate the response. Still, FOPT provides a 
meaningful leading order description of the dynamics (i.e. having no transitions), and therefore 
we do not regard the adiabatic non-linear regime as ``non-perturbative".

In the {\em non-perturbative regime} the evolution of $P_t(n|m)$ cannot be extracted from perturbation 
theory: not in leading order, neither in any order. Still it does not necessarily imply a non-linear
response. On the contrary: The semiclassical limit is contained in the deep non-perturbative 
regime \cite{CK00,CK03}. There, the LRT formula Eq.(\ref{e1}) is in fact valid. But its validity is 
{\em not} a consequence of perturbation theory, but rather the consequence of {\em quantal-classical
 correspondence} (QCC).

In the next subsection we will present a classical derivation 
of the general LRT expression (\ref{e1}). 
In Subsection \ref{sub:LRT-qm} we derive it using first order perturbation theory (FOPT). 
In Subsection \ref{sec:LRTSP} we derive the corresponding 
FOPT expression for the survival probability.

%%%%%%%%%%%%%%%%%%%%%%%%%%%%%%%%%%%%%%%%%%%%%%%%%%%%%%%%%%%%%%%%%%%%%
\subsection{Classical LRT derivation for $\delta E(t)$ \label{sub:LRT-cl}}

The classical evolution of  $E(t)={\mathcal{H}}(Q(t),P(t))$ 
can be derived from Hamilton equations. Namely, 
\be{0}
\frac{\dd{E}(t)}{\dd{t}}  
= [{\cal H},{\cal H}]_{\tbox{PB}} + \frac{\partial \mathcal{H}}{\partial t} 
= -\varepsilon \dot{f}(t) \mathcal{F}(t)\szk
\label{eq:LRT-dE1}
\ee
where $[ \cdot ]_{\rm PB}$ indicates the Poisson Brackets. Integration of Eq. (\ref{eq:LRT-dE1})
leads to 
\be{0}
E(t)-E(0) =  - \varepsilon \int_{0}^{t}  \mathcal{F}(t')\dot{f}(t')dt'\szp
\label{eq:LRT-dE2}
\ee
Taking a micro-canonical average over initial conditions 
we obtain the following expression for the variance 
\begin{equation}
\delta E^{2}(t)  
= \varepsilon^{2} \int_{0}^{t} C(t'-t'') \, 
\dot{f}(t')\dot{f}(t'') dt' dt''\szk
\label{eq:LRT-cl}
\end{equation}
which can be re-written in the form of (\ref{e1}).

One extreme special case of Eq.(\ref{e1}) is the sudden limit 
for which $f(t)$ is a step function. Such evolution
is equivalent to the LDoS studies of Section \ref{sec:regimes}. 
In this case $F_t(\omega)=1$, and accordingly
\be{0} 
\label{e2}
\delta E_{\tbox{cl}} \ = \  \varepsilon \times \sqrt{C(0)}
\hspace*{2cm} \mbox{\small [``sudden" case]}\szp
\ee

Another extreme special case is the response 
for persistent (either linear or periodic) driving 
of a system with an extremely short correlation time.  
In such case $F_t(\omega)$ becomes a narrow function 
with a weight that grows linearly in time. 
For linear driving ($f(t)=t$) 
we get $F_t(\omega) = t \times 2\pi\delta(\omega)$. 
This implies diffusive behavior:
\be{0}
\label{e3}
\delta E(t) = \sqrt{2 D_E t}
\hspace*{2cm} \mbox{\small [``Kubo" case]}\szk
\ee
where $D_{E} \propto \epsilon^2 $ is the diffusion coefficient.
The expression for $D_{E}$ as an integral over the correlation function 
is known in the corresponding literature either as Kubo formula, or as Einstein relation, 
and is the corner stone of the Fluctuation-Dissipation relation.

%%%%%%%%%%%%%%%%%%%%%%%%%%%%%%%%%%%%%%%%%%%%%%%%%%%%%%%%%%%%%%%%%%%%%
\subsection{Quantum LRT derivation for $\delta E(t)$\label{sub:LRT-qm}} 

The quantum mechanical derivation looks like an exercise 
in first order perturbation theory. In fact a proper 
derivation that extends and clarifies the regime where 
the result is applicable requires infinite order. 
If we want to keep a complete analogy with the classical 
derivation we should work in the adiabatic basis \cite{C00}. 
(For a brief derivation see Appendix D of \cite{WC02}).

In the following presentation we work in a "fixed basis" 
and assume $f(t)=f(0)=0$. We use the 
standard textbook FOPT expression for the transition
probability from an initial state $m$ 
to any other state $n$. This is followed by integration 
by parts. Namely, 
\begin{eqnarray}
\qquad\qquad P_{t}(n|m) & = & 
\frac{\varepsilon^{2}}{\hbar^{2}}|\bm{B}_{nm}|^{2}
\left|\int\limits _{0}^{t}\dd{t'}f(t')\eexp{i(E_{n}-E_{m})t'/\hbar}\right|^{2}\nonumber \\
& = & 
\frac{\varepsilon^{2}}{\hbar^{2}}|\bm{B}_{nm}|^{2}
\frac{ \tilde{F}_{t}(\omega_{nm}) } { (\omega_{nm})^2 } \szk
\label{eq:FOPT-ker}
\end{eqnarray}
where $\omega_{nm}=(E_n-E_m)/\hbar$. 
Now we calculate the variance and use Eq.~(\ref{eq:bandprofile}) 
so as to get 
\begin{eqnarray}
\qquad\qquad\qquad\delta E^{2}(t) & = & \sum_{n}P_{t}(n|m)(E_{n}-E_{m})^{2}\nonumber \\
 & = & \varepsilon^{2}\int_{-\infty}^{\infty}\frac{\dd{\omega}}{2\pi}\,\tilde{C}(\omega)\,
\tilde{F}_{t}(\omega)\szp
\label{eq:LRT-qm-var}
\end{eqnarray}

%%%%%%%%%%%%%%%%%%%%%%%%%%%%%%%%%%%%%%%%%%%%%%%%%%%%%%%%%%%%%%%%%%%%%
\subsection{Restricted QCC} 

The FOPT result for $\delta E(t)$ is \emph{exactly} the same 
as the classical expression Eq.~(\ref{e1}). 
It is important to realize that there is no $\hbar$-dependence 
in the above formula. This correspondence does not hold 
for the higher $k-$moments of the energy distribution. 
If we use the above FOPT procedure we get that the latter 
scale as $\hbar^{k-2}$.   

We call the quantum-classical correspondence for the second moment 
"restricted QCC". It is a very robust correspondence \cite{CK03}.
This should be contrasted with "detailed QCC" that applies 
only in the semiclassical regime where $P_{t}(n|m)$ can be 
approximated by a classical result (and not by a perturbative result).

%%%%%%%%%%%%%%%%%%%%%%%%%%%%%%%%%%%%%%%%%%%%%%%%%%%%%%%%%%%%%%%%%%%%%
\subsection{Quantum LRT derivation for $\mathcal{P}(t)$ \label{sec:LRTSP}}

With the validity of FOPT assumed we can also calculate  
the time-decay of the survival probability ${\cal P}(t)$. 
From Eq.~(\ref{eq:FOPT-ker}) we get:
\begin{equation}
p(t) \equiv \sum_{n(\ne n_0)} P_t(n|m) 
= \varepsilon^{2}\int_{-\infty}^{\infty} {d\omega\over 2\pi}
{\tilde C}(\omega) {{\tilde F}_t(\omega)\over(\hbar\omega)^2}\szp
\label{eq:FOPT-P}
\end{equation}
Assuming that $\mathcal{P}(t) = 1-p(t)$ 
can be extrapolated  in a "stochastic" fashion 
we get Eq.~(\ref{eq:LRT-qm-P}).
Another way to write the final formula is as follows:
\begin{equation}
\mathcal{P}(t) =  
\exp \left[ -\frac{1}{\hbar^2} 
\int_0^t \int_0^t  C(t'{-}t'') \delta x (t') \delta x(t'') dt'dt'' \right]\szp
\end{equation}
For constant perturbation (wavepacket dynamics) and assuming long times  
we obtain the Wigner decay, 
\be{0}
\label{wpptL}
{\cal P}(t) = \exp\left[- \left({\epsilon\over \hbar}\right)^2{\tilde C}(\omega{=}0) \times t\right]\szk
\ee
which can be regarded  
as a special case of Fermi-Golden-Rule.

%%%%%%%%%%%%%%%%%%%%%%%%%%%%%%%%%%%%%%%%%%%%%%%%%%%%%%%%%%%%%%%%%%%%%
\subsection{Note on $\mathcal{P}(t)$ for a time reversal scenario}

The "LRT formula" for $\mathcal{P}(t)$ 
in the case of "driving reversal scenario" is 
\begin{equation}
\mathcal{P}_{\tbox{DR}}(t) =  
\exp \left[ -\left(\frac{\varepsilon}{\hbar}\right)^2 
\int_0^T \int_0^T  C(t'{-}t'') f(t') f(t'') dt'dt'' \right]\szk
\end{equation}
where we assumed the simplest scenario with ${f(t)=1}$ for ${0<t<(T/2)}$ 
and  ${f(t)=-1}$ for ${(T/2)<t<T}$. 
It is interesting to make a comparison with the 
analogous result in case of "time reversal scenario".

The well known Feynman-Vernon influence functional 
has the following approximation: 
\begin{eqnarray}
\quad F[x_A,x_B] & = & \langle \Psi | U[x_B]^{-1} U[x_A] | \Psi \rangle  
\\ \nonumber   
            & = & \exp \left[ -\frac{1}{2\hbar^2} 
\int_0^t \int_0^t  C(t'{-}t'') (x_B(t'){-}x_A(t''))^2 dt'dt'' \right]\szp
\end{eqnarray}
This expression is in fact exact in the case of harmonic bath, 
and assuming thermal averaging over the initial state. 
Otherwise it should be regarded as an extrapolated 
version of leading order perturbation theory 
(as obtained in the interaction picture). 
What people call nowadays "fidelity" or ``Loschmidt echo" is in fact 
a special case of the above expression which is defined 
by setting $t=T/2$ and  $x_A=\varepsilon/2$ 
while $x_B=-\varepsilon/2$. Thus  
\begin{eqnarray}
 \qquad\qquad\mathcal{P}_{\tbox{TR}}(t)& = & | F[x_A,x_B] |^2  \\ \nonumber
 & = & \exp \left[ -\left(\frac{\varepsilon}{\hbar}\right)^2 
\int_0^{T/2} \int_0^{T/2}  C(t'{-}t'') dt'dt'' \right]\szp
\end{eqnarray}
Assuming a very short correlation time one obtains
\be{0}
{\cal P}_{\tbox{TR}}(T) = 
\exp\left[- \frac{1}{2}\left(\frac{\epsilon}{\hbar}\right)^2
{\tilde C}(\omega{=}0) \times T\right]\szk
\ee
which again can be regarded  
as a special variation of the Fermi-Golden-Rule 
(but note the pre-factor $1/2$).

%%%%%%%%%%%%%%%%%%%%%%%%%%%%%%%%%%%%%%%%%%%%%%%%%%%%%%%%%%%%%%%%%%%%%
\subsection{The survival probability and the LDoS}

For constant perturbation it is useful to remember that $\mathcal{P}(t)$ 
LDoS as follows:
\begin{eqnarray}
\qquad\qquad\qquad\mathcal{P}(t) & \equiv & \left|\langle n(x_{0})|e^{-i{\cal H}(x)t/\hbar}|n(x_{0})\rangle\right|^{2}\nonumber \\
 & = & \left|\sum_{m}e^{-i E_{m}(x) t/\hbar}| \langle m(x)|n(x_{0})\rangle|^{2}\right|^{2}\nonumber \\
 & = & \left| \int_{\infty}^{\infty}  P(E|m) \eexp{-iEt/\hbar}  dE \right|^{2}\szp
\label{eq:LDOS_is_FT_of_P}
\end{eqnarray}
This implies that a Wigner decay is associated with 
a Lorentzian approximation for the LDoS. 
In the non-perturbative regime the LDoS is not a Lorentzian, 
and therefore one should not expect an exponential.
In the semiclassical regime the LDoS shows 
system specific features and therefore the decay 
of $\mathcal{P}(t)$ becomes non-universal.

%%%%%%%%%%%%%%%%%%%%%%%%%%%%%%%%%%%%%%%%%%%%%%%%%%%%%%%%%%%%%%%%%%%%%%%
\section{Wavepacket Dynamics for Constant Perturbation \label{sec:WP-dyn}}

\begin{figure}[t]
\centerline{\includegraphics[width=1\hsize,angle=0,clip]{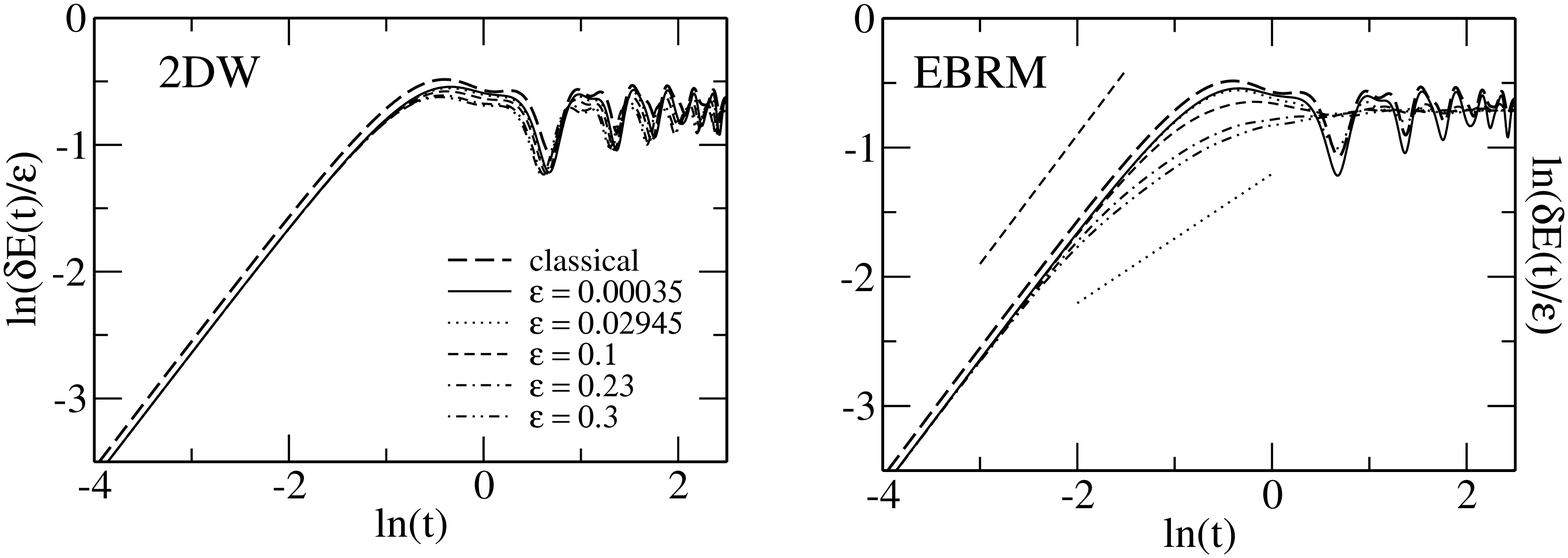}}
\caption{\label{cap:dE-2DW}
Simulations of wavepacket dynamics for the 2DW model (left panel) and for the corresponding 
EBRM model (right panel). The 
energy spreading $\delta E(t)$ 
(normalized with respect to the perturbation strength $\epsilon$) 
is plotted as a function of 
time for various perturbation strengths $\epsilon$ corresponding to different line-types
(the same in both panels). The classical spreading $\delta E_{\rm cl}(t)$ (thick dashed 
line) is plotted in both panels as a reference.}
\end{figure}

The first evolution scheme that we are investigating here is the so-called \emph{wavepacket 
dynamics}. The classical picture is quite clear \cite{CIK00,KC01}: 
The initial preparation is assumed to be a micro-canonical distribution that is supported by the 
energy surface ${\mathcal{H}}_{0} (Q,P)=E(0)$. Taking ${\mathcal{H}}$ to be a generator for the 
classical dynamics, the phase-space distribution spreads away from the initial surface for $t>0$. 
`Points' of the evolving distribution move upon the energy surfaces of ${\mathcal{H}}(Q,P)$. 
Thus, the energy $E(t)={\mathcal{H}}_{0}(Q(t),P(t))$ of the evolving distributions spreads 
with time.  Using the LRT formula Eq.(\ref{eq:spectral-content-F})
for rectangular pulse $f(t')=1$ for $0< t'<t$ we get 
\begin{equation}
\tilde{F}_{t}(\omega)=\left|1-\eexp{-i\omega t}\right|^{2} = (\omega t)^2 {\rm sinc}
\left({\omega t\over 2}\right)\szk
\label{eq:spec_wpk}
\end{equation}
and hence
\begin{equation}
\delta E_{\tbox{cl}}(t)=\varepsilon\times\sqrt{2(C(0)-C(t))}\szp
\label{eq:LRT-dE-wpk}
\end{equation}
For short times $t\ll \tau_{\tbox{cl}}$ we can expand the correlation function as 
$C(t)\approx C(0)-{1\over2}C''(0)t^2$, leading to a ballistic evolution.
Then, for $t\gg\tau_{\tbox{cl}}$, due to ergodicity, a `steady-state distribution' appears, 
where the evolving `points' occupy an `energy shell' in phase-space. The thickness of this energy 
shell equals $\delta E_{\tbox{cl}}$. Thus, we have a crossover from ballistic energy spreading 
to saturation:
\begin{equation}
\delta E(t)\approx\left\{ 
\begin{array}{lcr}
\sqrt{2}(\ecl/\tau_{\tbox{cl}})\, t & \mbox{for} & t<\tau_{\tbox{cl}}\\
\sqrt{2} \ecl & \mbox{for} & t>\tau_{\tbox{cl}}
\end{array}\right.\szp
\label{eq:LRT-dE-wpk2}
\end{equation}
Figure \ref{cap:dE-2DW} shows the classical energy spreading (heavy dashed line) for the 2DW model. In agreement 
with Eq.~(\ref{eq:LRT-dE-wpk2}) we see that $\delta E_{\tbox{cl}}(t)$ is first ballistic 
and then saturates. The classical dynamics is fully characterized by 
the two classical parameters $\tau_{\tbox{cl}}$ and $\delta E_{\tbox{cl}}$.

%%%%  FIGURE %%%%%%%%%%%%%%%%%%%%%%%%%%
%%%% END FIGURE %%%%%%%%%%%%%%%%%%%%%%%

%%%%%%%%%%%%%%%%
\subsection{The quantum dynamics}

Let us now look at the quantized 2DW model. The quantum mechanical data are reported in Fig. 
\ref{cap:dE-2DW} (left panel) where different curves correspond to various perturbation strengths $\varepsilon$.
As in the classical case (heavy dashed-line) we observe an initial ballistic-like spreading 
\cite{KC01} followed by saturation. This could lead to the wrong impression that the classical 
and the quantum spreading are of the same nature. However, this is definitely not the case.

\begin{figure}[t]
\centerline{\includegraphics[width=0.5\hsize,angle=0,clip]{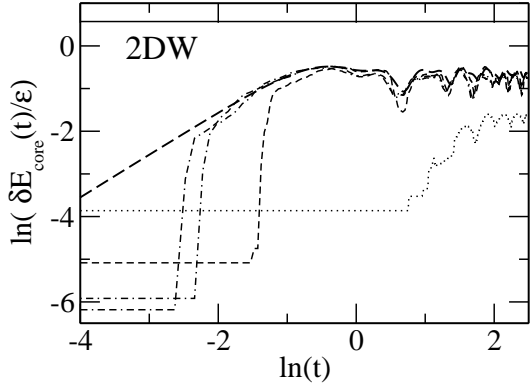}}
\caption{\label{cap:N-2DW}
Simulations of wavepacket dynamics for the 2DW model. 
The evolution of the (normalized) core width $\delta E_{\rm core}(t)$ 
is plotted as a function of time. The classical 
expectation is represented by a thick dashed line 
for the sake of comparison. As $\epsilon$ becomes larger
it is approached more and more. We use the same set of parameters
as in Fig.~\ref{cap:dE-2DW}.}
\end{figure}

In order to detect the different nature of quantum ballistic-like spreading, one has to inquire 
measures that are sensitive to the structure of the profile, such as 
the core-width $\delta E_{\tbox{core}}(t)$.  
In Fig.~\ref{cap:N-2DW} we present our numerical data for the 2DW model. If the spreading were 
of a classical type, it would imply that the spreading profile is characterized 
by a single energy scale. In such a case we would expect that $\delta E_{\tbox{core}}(t) \sim \delta E(t)$. 
Indeed this is the case for $\varepsilon >\varepsilon_{\tbox{prt}}$ with the exclusion of 
very short times: The larger $\varepsilon$ is the shorter the quantal transient becomes.
In the perturbative regimes, in contrast to the semiclassical regime,  
we have a separation of energy scales $\delta E_{\tbox{core}}(t) \ll \delta E(t)$. 
In the perturbative regimes $\delta E(t)$ is determined by the tails, 
and it is not sensitive to the size of the `core' region.  

%%%%  FIGURE %%%%%%%%%%%%%%%%%%%%%%%%%%
%%%% END FIGURE %%%%%%%%%%%%%%%%%%%%%%%

Using the LRT formula for $\mathcal{P}(t)$ we get, for 
short times  ($t \ll \tau_{\tbox{cl}}$) during the ballistic-like stage
\be{0}
\label{wpptS}
{\cal P}(t) = {\rm exp}\left(-C(\tau{=}0)\times \left({\epsilon t\over \hbar}\right)^2\right)\szk
\ee
while for long times  ($t \gg \tau_{\tbox{cl}}$) we have 
the FGR decay of Eq.(\ref{wpptL}).
Can we trust these expressions? 
Obviously FOPT can be trusted  
as long as $\mathcal{P}(t)\sim 1$.  
This can be converted into an inequality $t < t_{\tbox{prt}}$ where 
\begin{equation}
t_{\tbox{prt}} = \left(\frac{\varepsilon_{\tbox{prt}}}{\varepsilon}\right)^{\nu=1,2}\tau_{\tbox{cl}}\szp
\label{eq:tprt}
\end{equation}
The power $\nu=1$ applies to the non-perturbative regime where the breakdown of 
$\mathcal{P}(t)$ happens to be before $\tau_{\tbox{cl}}$. The power $\nu=2$ applies 
to the perturbative regime where the breakdown of $\mathcal{P}(t)$  happens 
after $\tau_{\rm cl}$ at $t_{\tbox{prt}}=\hbar/\Gamma$, i.e. after the ballistic-like 
stage.

The long term behavior of  $\mathcal{P}(t)$ in the non-perturbative 
regime is not the Wigner decay. It can be obtained by Fourier 
transform of the LDoS. In the non-perturbative regime the LDoS 
is characterized by the single energy scale $\delta E_{\tbox{cl}}\propto \delta x$. 
Hence the decay in this regime is characterized by 
a semiclassical time scale $2\pi\hbar / \delta E_{\tbox{cl}}$.

%%%%%%%%%%%%%%%%
\subsection{The EBRM dynamics \label{sec:WP-dynRR}}

Next we investigate the applicability of the RMT approach to describe wavepacket dynamics 
\cite{CIK00,KC01} and specifically the energy spreading $\delta E(t)$. 
At first glance, we might be tempted to speculate that RMT should be able, 
at least as far as $\delta E(t)$ is concerned, to describe the actual quantum picture. 
After all, we have seen in Subsection \ref{sec:LRT} that the quantum mechanical 
LRT formula (\ref{eq:LRT-qm-var}) for the energy spreading involves as its only input 
the classical power spectrum ${\tilde C}(\omega)$. Thus we would expect that an effective RMT model 
with the same band-profile would lead to the \emph{same} $\delta E(t)$.

However, things are not so trivial. In Figure \ref{cap:dE-2DW} we show the numerical results 
for the EBRM model \footnote{The same qualitative results were found also for the prototype
WBRM model, see \cite{CIK00}.}. In the standard and in the extended perturbative regimes 
we observe a good agreement with Eq.(\ref{eq:LRT-qm-var}). 
This is not surprising as the theoretical prediction was derived via FOPT, 
where correlations between off-diagonal elements are not important. 
In this sense the equivalence of the 2DW model and the EBRM model 
is trivial in these regimes. But as soon as we enter the non-perturbative regime, 
the spreading $\delta E(t)$ shows a qualitatively different behavior 
from the one predicted by LRT: After an initial ballistic spreading, we observe 
a premature crossover to a diffusive behavior 
\begin{equation}
\delta E(t)=\sqrt{2D_{E}t}\szp
\label{eq:D:diffusion-WEBRM}
\end{equation}

The origin of the diffusive behavior can be understood in the following way. Up to time 
$t_{\tbox{prt}}$ the spreading $\delta E(t)$ is described accurately by the FOPT result 
(\ref{eq:LRT-qm-var}). At $t\sim t_{\tbox{prt}}$ the evolving distribution becomes 
as wide as the bandwidth, and we have $\delta E_{\tbox{core}} \sim \delta E \sim \Delta_b$ 
rather than $\delta E_{\tbox{core}} \ll \delta E \ll \Delta_b$. 
We recall that in the non-perturbative regime FOPT is subjected to a breakdown 
before reaching saturation. The following simple heuristic picture turns out to be correct. 
Namely, once the mechanism for ballistic-like spreading disappears, 
a stochastic-like behavior takes its place. 
The stochastic energy spreading is similar 
to a random-walk process where the step size 
is of the order $\Delta_{b}$, with transient 
time $t_{\tbox{prt}}$. Therefore we have 
a diffusive behavior $\delta E(t)^{2}=2D_{\tbox{E}}t$ with 
\be{0}
D_{\tbox{E}}\,\,=\,\, C\cdot\Delta_{b}^{2}/t_{\tbox{prt}}\,\,=C\cdot\Delta^{2}b^{\rm 5/2}
\varepsilon\sigma/\hbar\,\,\,\,\propto\,\,\,\,\hbar\label{eq:D_E-WP}
\ee
where $C$ is some numerical pre-factor. This diffusion is not of classical nature, since in the
$\hbar\rightarrow 0$ limit we get $D_E\rightarrow 0$. The diffusion can go on until the energy
spreading profile ergodically covers the whole energy shell and saturates to a classical-like 
steady state distribution. The time $t_{\tbox{erg}}$ for which we get ergodization is characterized
by the condition $(D_{\tbox{E}}t)^{\rm 1/2}<\delta E_{\tbox{cl}}$, leading to
\be{0}
\label{tergn1}
t_{\tbox{erg}}\,\,=\,\, b^{\rm -3/2}\,\,\hbar\,\varepsilon\,\sigma/\Delta^{2}\,\,\propto\,\,1/\hbar\szp
\ee

\begin{figure}[t]
\includegraphics[width=1\hsize,angle=0,clip]{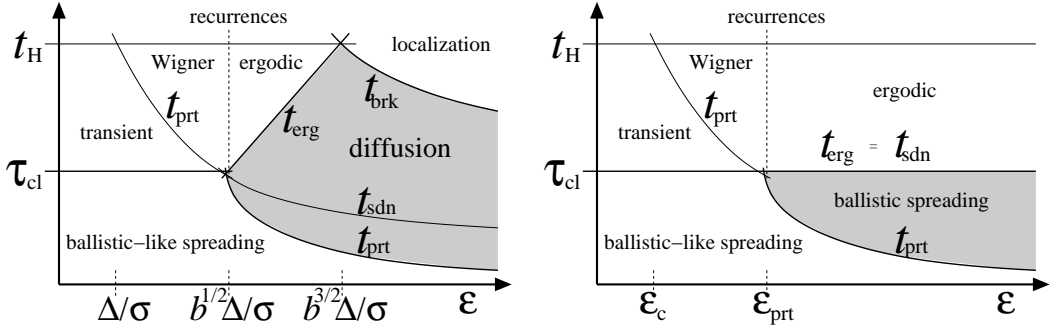}
\caption{\label{brm_diag}
A diagram that illustrates 
the various time scales in wavepacket dynamics, 
depending on the strength of the perturbation $\varepsilon$. 
The diagram on the left refers to the WBRM model, 
while that on the right is for a quantized system that 
has a classical limit. The two cases differ in 
the non-perturbative regime (large  $\varepsilon$): 
In the case of a quantized model we have a genuine 
ballistic behavior which reflects detailed QCC, 
while in the RMT case we have a diffusive stage.
In the latter case the times scale $t_{\tbox{sdn}}$ 
marks the crossover from reversible to non-reversible 
diffusion. This time scale can be detected 
in a driving reversal scenario as explained in the next section. 
For further discussion of this diagram see the text, 
and in particular the concluding section of this paper.}  
\end{figure}

For completeness we note that for $\varepsilon>\varepsilon_{loc}$ 
there is no ergodization but rather dynamical ("Anderson" type) localization. 
Hence, in the latter case, $t_{\tbox{erg}}$ is replaced by the break-time  $t_{\tbox{brk}}$. 
The various regimes and time scales are illustrated by the diagram presented in Fig. \ref{brm_diag}.

%%%%  FIGURE %%%%%%%%%%%%%%%%%%%%%%%%%%
%%%% END FIGURE %%%%%%%%%%%%%%%%%%%%%%%

%%%%%%%%%%%%%%%%%%%%%%%%%%%%%%%%%%%%%%%%%%%%%%%%%%%%%%%%%%%%%
\section{Driving Reversal Scenario \label{sec:DR}}

A thorough understanding of the one-period driving reversal scenario \cite{KC03} is both 
important within itself, and for constituting a bridge towards a theory dealing with the response 
to periodic driving \cite{CK00}. In the following subsection we present our results 
for the prototype WBRM model, 
while in Subsection \ref{sub:DR-2DW} we consider the 2DW model 
and compare it to the corresponding EBRM model.
The EBRM is better for the purpose of making comparisons 
with the 2DW, while the WBRM is better for the sake of quantitative analysis 
(the "physics" of the EBRM and the WBRM models is, of course, the same).

\begin{figure}[t]
\begin{center}
\includegraphics[width=1\hsize,angle=0,clip]{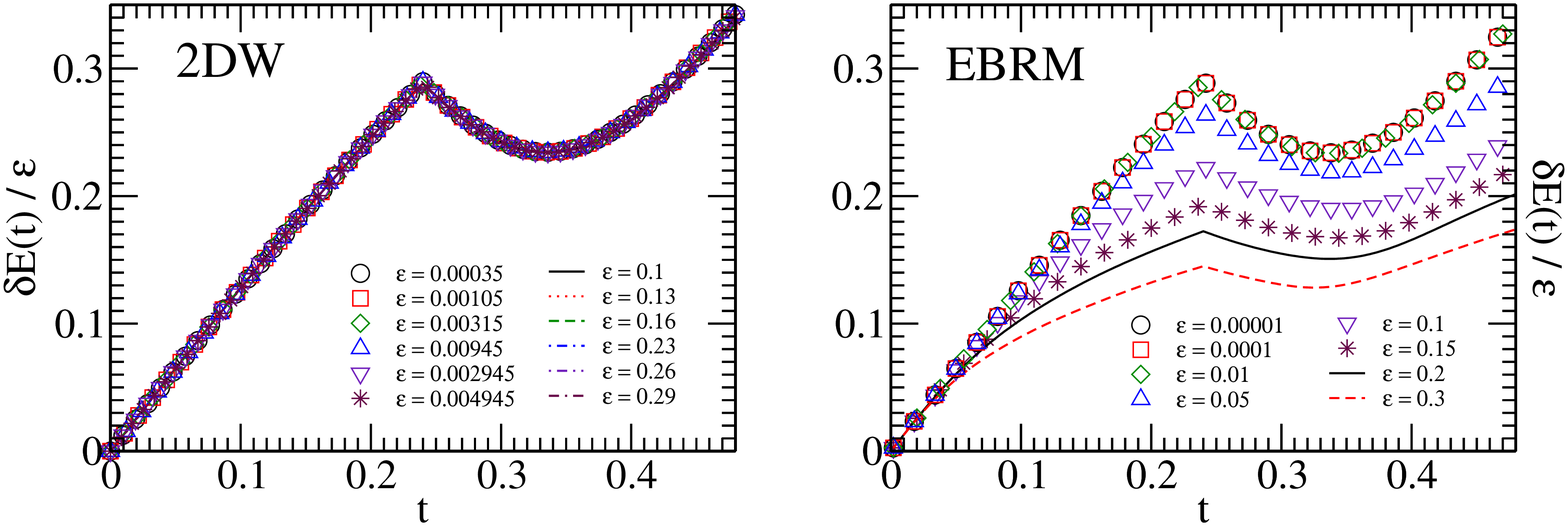}
\includegraphics[width=1\hsize,angle=0,clip]{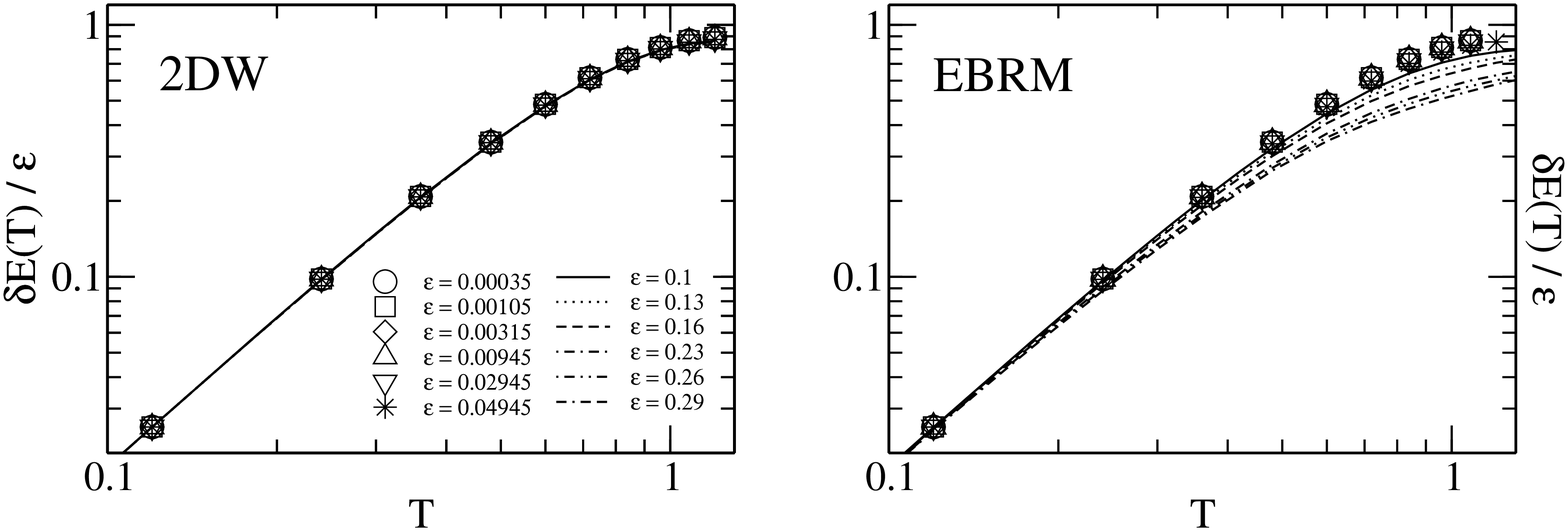}
\end{center}

\caption{\label{DRde}
Simulations of driving reversal for the 2DW model (left panels) and for the corresponding 
EBRM model (right panels). In the upper row the (normalized) energy spreading $\delta E(t)$ 
is plotted as a function of time for representative values $\varepsilon$ 
while $T=0.48$. In the lower row the  
(normalized) energy spreading $\delta E(T)$ at the end of the cycle 
is plotted versus $T$ for representative values of $\varepsilon$.} 
\end{figure}

\begin{figure}[t]
\begin{center}
\includegraphics[width=1\hsize,angle=0,clip]{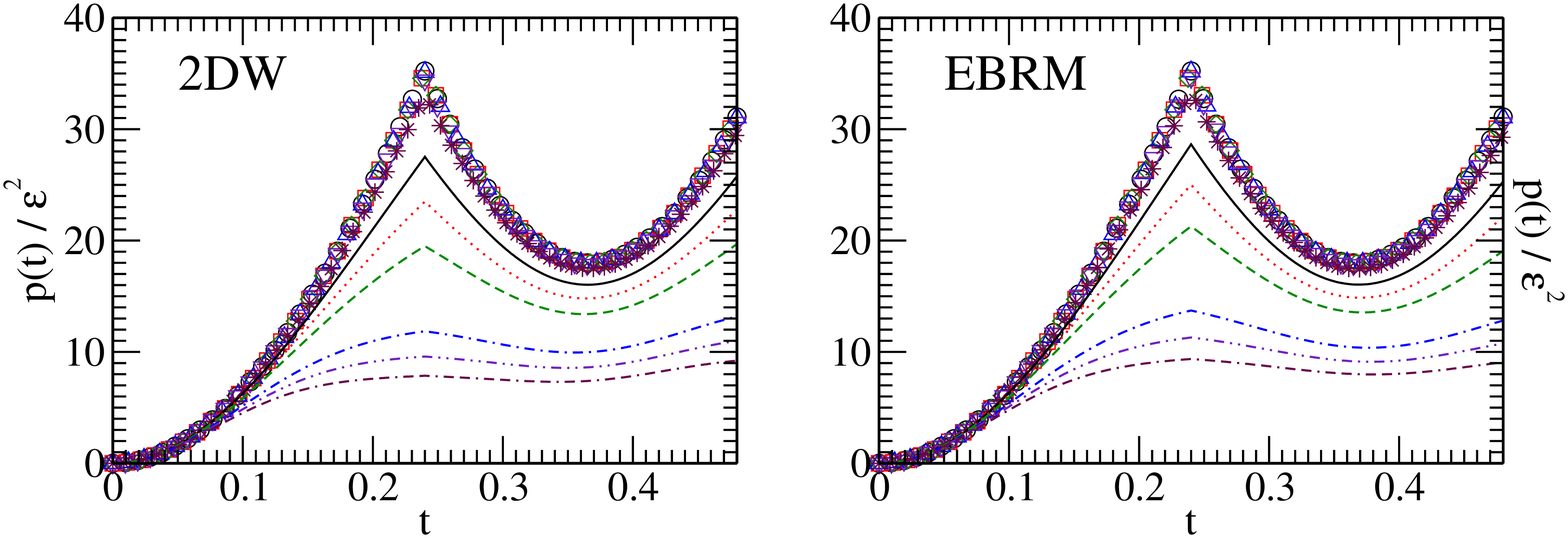}
\includegraphics[width=1\hsize,angle=0,clip]{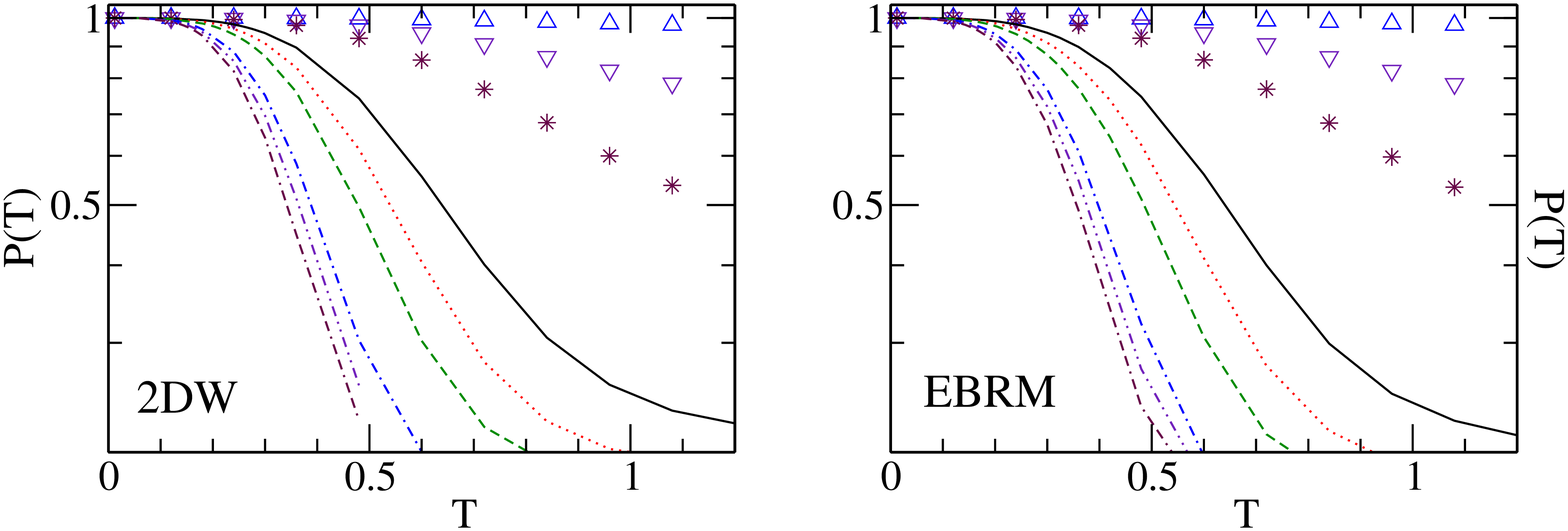}
\end{center}
\caption{\label{DRpt}
Simulations of driving reversal for the 2DW model (left panels) and for the corresponding 
EBRM model (right panels). In the upper row the (normalized) transition probability $p(t)=1-
\mathcal{P}(t)$ is plotted as a function of time for representative values $\varepsilon$. 
The period of the driving is $T=0.48$. In the lower row the survival probability $\mathcal
{P}(T)$ at the end of the cycle is plotted versus $T$ for representative values of 
$\varepsilon$. In all cases we are using the same symbols as in the upper left panel of 
Fig. \ref{DRde}.}
\end{figure}

The quantities that monopolize our interest are 
the energy spreading $\delta E(t)$ and the 
survival probability  ${\cal P}(t)$.  
In Figs. \ref{DRde} and \ref{DRpt} we present representative plots. 
From a large collection of such data
that collectively span a very wide range 
of parameters,  we extract results 
for $\delta E(T)$, for ${\cal P}(T)$,  
and for the corresponding compensation times. 
These are presented in Figs.~\ref{DRde},\ref{DRpt},\ref{DRtrE},\ref{DRPT},\ref{DRtrP} and \ref{tP}.

%%%%  FIGURE %%%%%%%%%%%%%%%%%%%%%%%%%%
%%%% END FIGURE %%%%%%%%%%%%%%%%%%%%%%%

%%%%  FIGURE %%%%%%%%%%%%%%%%%%%%%%%%%%

%%%% END FIGURE %%%%%%%%%%%%%%%%%%%%%%%

%%%%%%%%%%%%%%%%%%%%%%%%%%%%%%%%%%%
\subsection{Driving Reversal Scenario: RMT Case \label{sub:DR-WBRM}}

\subsubsection{LRT for the energy spreading}

Assuming that the driving reversal happens    
at $t=T/2$, the spectral content ${\tilde F}_t(\omega)$
for $T/2<t<T$ is
\begin{equation}
\tilde{F}_{t}(\omega)=\left|1-2\eexp{-i\omega T/2}+\eexp{i\omega t}\right|^{2}\szp
\label{eq:spec_DR}
\end{equation}
Inserting Eq.~(\ref{eq:spec_DR}) into Eq.~(\ref{e1}) we get
\begin{equation}
\label{eq:LRT-dE-WBRMa}
\delta E(t) = \varepsilon \times \sqrt{6C(0)+2C(t)-4C({T\over 2})-4C(t{-}{T\over2})}\szp
\end{equation}
For the WBRM model we can substitute in Eq.~(\ref{eq:LRT-dE-WBRMa}) 
the exact expression Eq.~(\ref{wbrmct}) for the correlation function, and get
\begin{equation}
\label{wbrmdEt}
\delta E(t) =  2 \varepsilon \sigma \times \sqrt{3 b + b {\rm sinc} ({t\over t_{\rm cl}})-
2 b {\rm sinc} ({T\over 2 \tau_{\rm cl}})-2 b {\rm sinc} ({t{-}{T\over 2}\over \tau_{\rm cl}})}\szp
\end{equation}
We can also find the compensation time $t_r^E$ by minimizing Eq.~(\ref{eq:LRT-dE-WBRMa})
with respect to $t$. For the WBRM model we have 
\begin{equation}
{2\cos\!\left[\frac{T/2-t}{\tau_{\tbox{cl}}}\right]\!\over \tau_{\rm cl} (T/2-t)}+
{2\sin\!\left[\frac{T/2-t}{\tau_{\tbox{cl}}}\right]\!\over (T/2-t)^2}+
{\cos\!\left[\frac{t}{\tau_{\tbox{cl}}}\right]\over t\tau_{\rm cl}}=
\frac{1}{t^2}\sin\!\left[\frac{t}{\tau_{\tbox{cl}}}\right]\szk
\label{eq:LRT-t_r^M}
\end{equation}
which can be solved numerically to get $t_r^E$.

The spreading width at the end of the period is 
\begin{equation}
\label{eq:LRT-dE-WBRM}
\delta E(T) = \varepsilon \times \sqrt{6C(0)+2C(T)-8C({T\over2}))}\szp
\end{equation}

It is important to realize that the dimensional parameters 
in this LRT analysis are determined by the 
time scale $\tau_{\tbox{cl}}$ and by the energy scale $\delta E_{\tbox{cl}}$. 
This means that we have a scaling relation 
(using units such that $\sigma=\Delta=\hbar=1$)
\be{0}
\frac{\delta E(T)} {\sqrt{b} \, \varepsilon} 
=  h^E_{\tbox{LRT}} \left( b T \right) \szp
\ee
Deviation from this scaling relation implies 
a non-perturbative effect that goes beyond LRT.

The LRT scaling is verified nicely by our numerical data 
(see upper panels of Fig.~\ref{DRtrE} ). The values of
perturbation strength for which the LRT results are applicable 
correspond to $\varepsilon <\varepsilon_{\tbox{prt}}$. 
In the same Figure we also plot the whole analytical expression 
(\ref{eq:LRT-dE-WBRM}) for the spreading $\delta E(T)$. 
Similarly in Fig.~\ref{DRtrE} (lower panels) we present 
our results for the compensation time $t_{r}^{E}$.
All the data fall one on top of the other once we rescale them.
It is important to realize that the LRT scaling relation 
implies that the compensation time $t_{r}^{E}$ is \emph{independent} 
of the perturbation strength $\varepsilon$. It is determined 
only by the \emph{classical} correlation time $\cl{\tau}$. 
In the same figure we also present the resulting analytical 
result (heavy-dashed line) which had been obtained via Eq.(\ref{eq:LRT-t_r^M}). 
An excellent agreement with our data is evident.

%%%%%%%%%%%%%%%%%%%%%%%%%%%%%%%%%%%
\subsubsection{Energy spreading in the non-perturbative regime}

We turn now to discuss the dynamics in the non-perturbative
regime, which is our main interest. In the absence of driving
reversal (see Subsection \ref{sec:WP-dynRR}) we obtain diffusion 
(\mbox{$\delta E(t)\propto \sqrt{t}$}) for $t>t_{\tbox{prt}}$, where
\begin{equation}
\label{e5}
t_{\tbox{prt}} = \hbar/(\sqrt{b}\sigma\varepsilon)\quad .
\end{equation}
If $(T/2) < t_{\tbox{prt}}$, this non-perturbative diffusion
does not have a chance to develop, and therefore we can still trust Eq.~(\ref{eq:LRT-dE-WBRMa}).
So the interesting case is $(T/2) > t_{\tbox{prt}}$,
which means large enough $\varepsilon$.
In the following analysis we distinguish
between two stages in the non-perturbative diffusion process.
The first stage ($t_{\tbox{prt}}<t<t_{\tbox{sdn}}$)
is reversible, while the second stage ($t>t_{\tbox{sdn}}$) is irreversible.
For much longer time scales we have recurrences 
or localization, which are not the issue of this paper.
The new time scale ($t_{\tbox{sdn}}$)  did not appear in 
our "wavepacket dynamics" study, because it can be detected 
only by time driving reversal experiment.

The determination of the time scale $t_{\tbox{sdn}}$ 
is as follows. The diffusion coefficient is
$D_{\tbox{E}}=\Delta^2 b^{\tbox{5/2}} \sigma\varepsilon/\hbar$
up to a numerical pre-factor. The diffusion law 
is $\delta E^2(t)=D_{\tbox{E}}t$.
The diffusion process is reversible as long as
$\mbf{E}$ does not affect the relative phases of
the participating energy levels. This means that
the condition for reversibility is
$(\delta E(t) \times t) / \hbar \ll 1$.
The latter inequality can be written
as $t< t_{\tbox{sdn}}$, where
\begin{equation}
\label{e6}
t_{\tbox{sdn}} = \left(\frac{\hbar^2}{D_ {\tbox{E}}}\right)^{1/3} =
\left(\frac{\hbar^3}{\Delta^2b^{5/2}\sigma\varepsilon}\right)^{1/3}\szp
\end{equation}
It is extremely important to realize that without reversing the driving,
the presence or the absence of $\mbf{E}$ in the Hamiltonian cannot be detected.
It is only by driving reversal that we can easily determine (as in the
upper panels of Fig.\ref{DRde}) whether the diffusion process is reversible 
or irreversible.

\begin{figure}[t]
\begin{center}
\includegraphics[width=1\hsize,angle=0,clip]{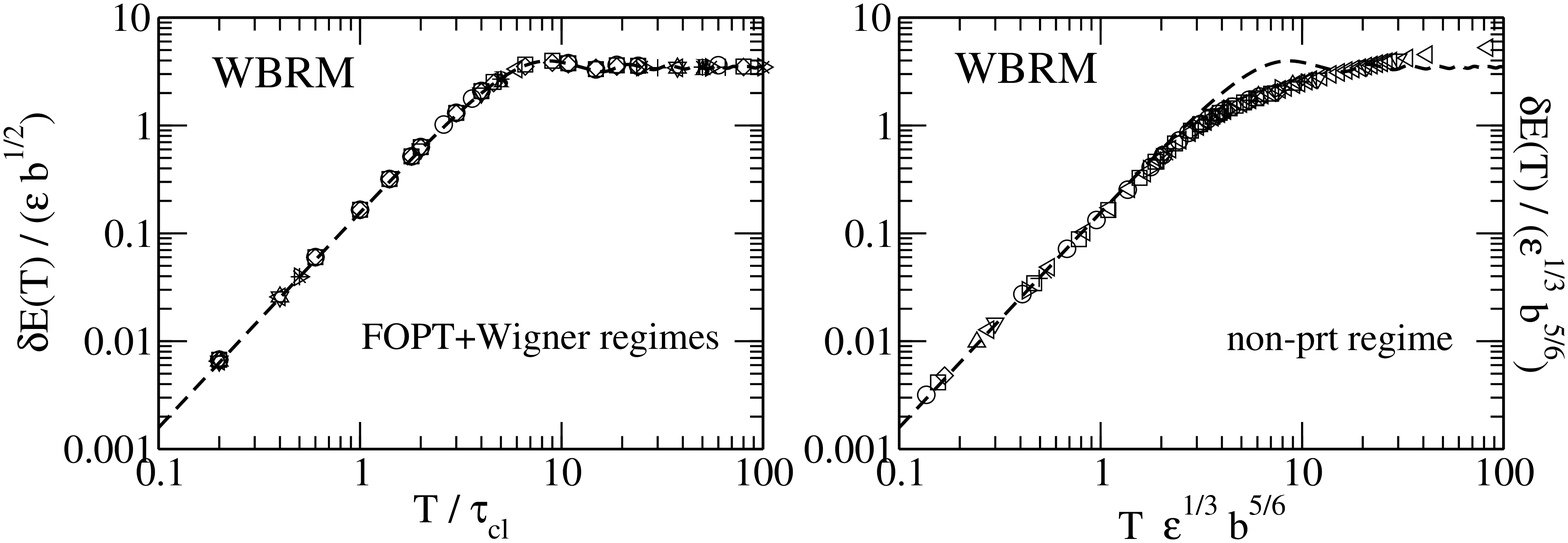}
\includegraphics[width=1\hsize,angle=0,clip]{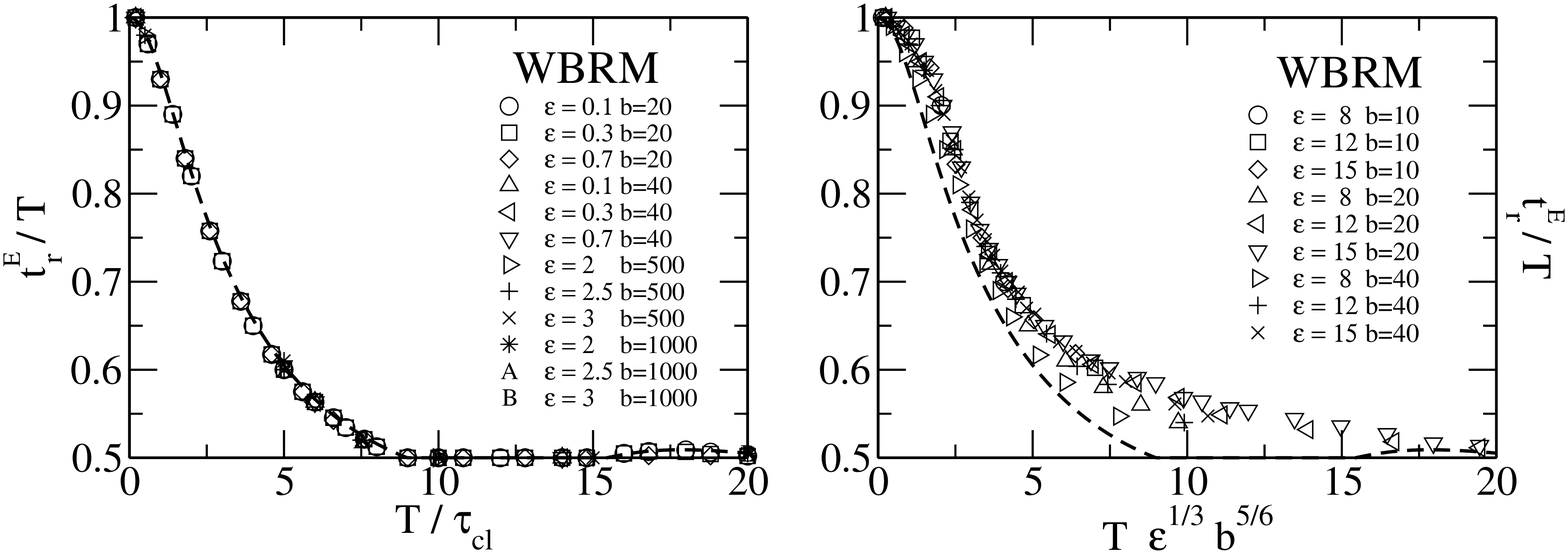}
\end{center}
\caption{\label{DRtrE}
Simulations of driving reversal for the WBRM model. 
In the upper row the (scaled) energy spreading $\delta E(T)$ 
at the end of the cycle is plotted against the (scaled) period $T$.  
In the lower panels the compensation time is plotted 
against the (scaled) period $T$. The panels on the left are for 
$\varepsilon$ values within the perturbative regime, 
while the panels on the left are for the non-perturbative 
regime. For the sake of comparison we plot 
the LRT expectation for $b=1$ as a heavy dashed line. } 
\end{figure}

The dimensional parameters in this analysis are 
naturally the time scale $t_{\tbox{sdn}}$ and the 
resolved energy scale $\hbar/T$. Therefore we 
expect to have instead of the LRT scaling, 
a different "non-perturbative" scaling relation. 
Namely, $\delta E(T) / (\hbar/T)$ 
should be related by a scaling function 
to $T/t_{\tbox{sdn}}$. Equivalently (using units such that \linebreak 
${\sigma=\Delta=\hbar=1}$) it can be written as 
\be{0}
\frac{\delta E(T)} { b^{5/6}\varepsilon^{1/3} } 
= h^E_{\tbox{nprt}} \left( b^{5/6} \varepsilon^{1/3}  T  \right)\szp
\ee
Obviously the non-perturbative scaling with 
respect to $\epsilon^{1/3}$ goes beyond any 
implications of perturbation theory. 
It is well verified by our numerical data 
(see upper right panel of Fig.~\ref{DRtrE}). 
The values of perturbation strength for which 
this scaling applies correspond 
to $\varepsilon > \varepsilon_{\tbox{prt}}$. 
The existence of the $t_{\tbox{sdn}}$ scaling 
can also be verified in the lower right panel of 
Fig.~\ref{DRtrE}, where we show that $t_r/T$ is 
by a scaling function related to $b^{5/6}\varepsilon^{1/3}T$.

%%%%  FIGURE %%%%%%%%%%%%%%%%%%%%%%%%%%
%%%% END FIGURE %%%%%%%%%%%%%%%%%%%%%%%

%%%%%%%%%%%%%%%%%%%%%%%%%%%%%%%%%%%
\subsubsection{Decay of $\mathcal{P}(t)$ in the FOPT regime}

We can substitute Eq.~(\ref{eq:spec_DR}) for the spectral 
content $\tilde{F}_{t}(\omega)$ 
of the driving into the LRT formula Eq.~(\ref{eq:FOPT-P}),  
and come out with the following expression 
for the survival probability at the end of the period $t=T$
\begin{equation}
{\cal P}(T) \approx \exp(-\varepsilon^{2}T^{4}b^{3})\szp
\label{eq:superexp}
\end{equation}
This is a super-Gaussian decay, which is quite different from 
the standard Gaussian decay Eq.~(\ref{wpptS}) or any other results on reversibility 
that appear in literature \cite{PS02,BC02,JP01,JAB02,HKCG04}.
We have verified that this expression is valid in the FOPT regime. See Figure \ref{DRPT}a.

For the WBRM model, we get the following expression for $p(t)$ after substituting 
the spectral content ${\tilde F}_t(\omega)$ given from Eq.~(\ref{eq:spec_DR})
\begin{equation}
\label{eq:tr1min}
p(t)= {(\epsilon\sigma)^2\over \Delta \hbar}
 \times \int_{-\omega_{\rm cl}}^{\omega_{\rm cl}} d\omega {6-4[\cos({\omega T\over 2})+\cos(\omega({T\over 2} -t))]+2cos(\omega t)\over \omega^2}\szp 
\end{equation}

The corresponding compensation time $t_r^P$ can be found after minimizing the 
above expression (corresponding to the maximization of $\mathcal{P}(t)=1-p(t)$) with 
respect to time $t$. This results in the following equation
\begin{equation}
\mbox{ si}\left(\omega_{cl}t\right)=2\mbox{ si}
\left(\omega_{cl}\left(t-\frac{T}{2}\right)\right)\szk
\label{eq:LRT-t_r^P}
\end{equation}
which has to be solved numerically in order to evaluate $t_r^P$. Above
$\mbox{si}(x)=\int_{0}^{x}\frac{\sin x}{x}$. Our numerical data are
reported in Fig.~\ref{DRtrP} together with the theoretical prediction (\ref{eq:LRT-t_r^P}).

%%%%%%%%%%%%%%%%%%%%%%%%%%%%%%%%%%%
\subsubsection{Decay of $\mathcal{P}(t)$ in the Wigner regime}

We now turn to discuss ${\cal P}(t)$ in the "Wigner regime". 
By this we mean $\varepsilon_c <  \varepsilon < \varepsilon_{\tbox{prt}}$.
This distinction does not appear in the $\delta E(t)$ analysis.
The time evolution of $\delta E(t)$ is dominated by the tails 
of the distribution and does not affect the "core" region. 
Therefore $\delta E(t)$ also agreed with LRT outside of the 
FOPT regime in the whole (extended) perturbative regime.   
But this is not the case with ${\cal P}(t)$, which is mainly 
influenced by the "core" dynamics.
As a result in the "Wigner regime" we get different 
behavior compared with the FOPT regime.

We look at the survival probability ${\cal P}(T)$
at the end of the driving period. In the Wigner regime, 
instead of the LRT-implied super-Gaussian decay, 
we find a Wigner-like decay:
\begin{equation}
\mathcal{P}(T) \approx \eexp{-\Gamma(\varepsilon) \, T}\szk
\label{eq:LRT-P(T)-wig}
\end{equation}
where  $\Gamma\approx\varepsilon^{2}/\Delta$. 
In Figure \ref{DRPT}b  we present 
our numerical results for various perturbation strengths
in this regime. A nice overlap is observed 
once we rescale the time axis as $\varepsilon^{2} \times T$. 
We would like to emphasize once more that both 
in the standard and in the extended perturbative regimes 
the scaling law involves the perturbation strength $\varepsilon$. 
This should be contrasted with the LRT scaling of $\delta E(t)$.

\begin{figure}[p]
\centerline{\includegraphics[height=0.85\vsize,angle=0,clip]{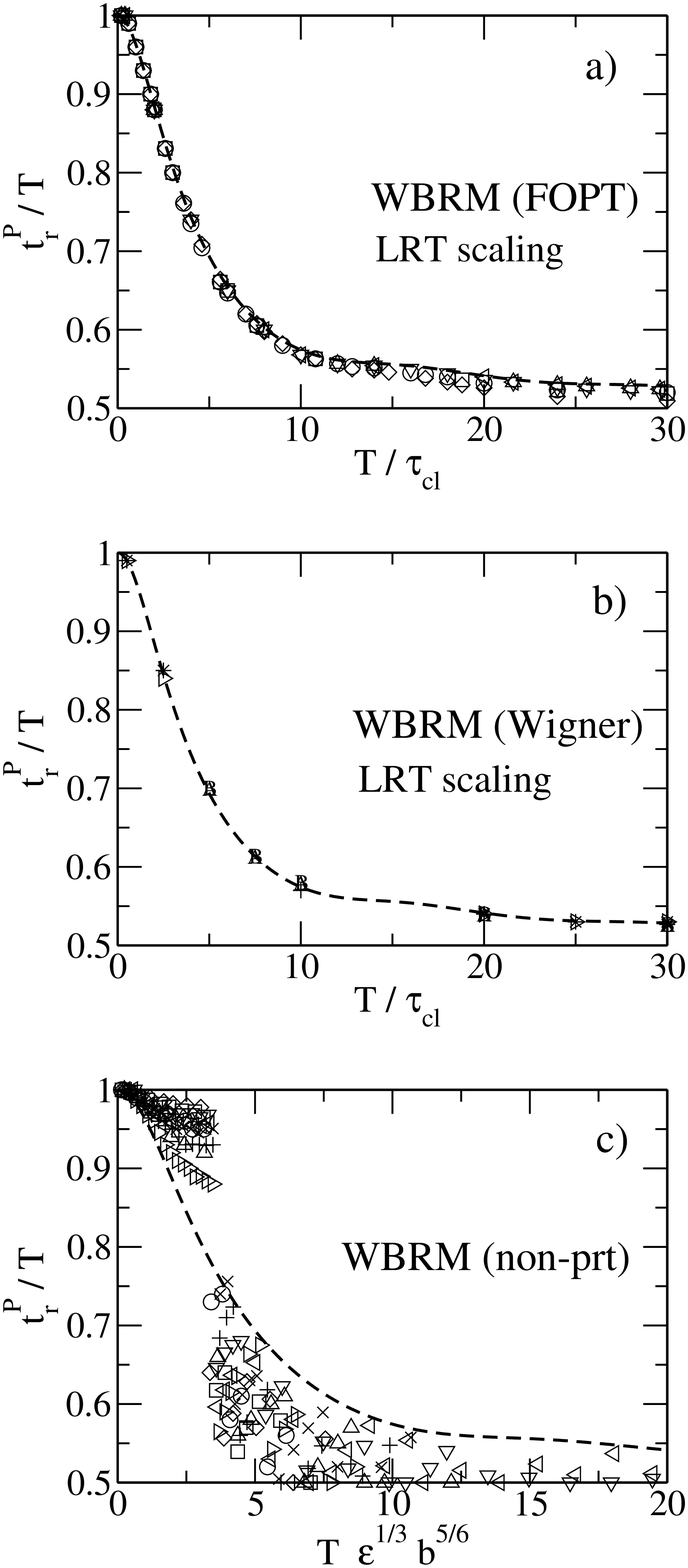}}
\caption{\label{DRtrP}
The compensation time is extracted for the same 
simulations of the previous figures. Note that 
for both the FOPT regime and the Wigner regime 
we have "LRT scaling". The thick dashed line corresponds to
the LRT prediction Eq.~(\ref{eq:LRT-t_r^P}) for $b=1$. Various 
symbols correspond to different ($\epsilon$,b) values such that
the $\epsilon < \epsilon_c$ in (a); $\epsilon_c <\epsilon < \epsilon_{\rm prt}$
in (b) and $\epsilon >\epsilon_{\rm prt}$ in (c). }
\end{figure}

What about the compensation time $t_r^P$? A reasonable assumption is that it will 
exhibit a different scaling in the FOPT regime and in the Wigner regime (as is 
the case of $\mathcal{P}(t)$). 
Namely, in the FOPT regime we would expect "LRT scaling" 
with $\tau_{\tbox{cl}}$, while in the Wigner regime 
we would expect "Wigner scaling" with $t_{\tbox{prt}}=\hbar/\Gamma$. 
The latter is of non-perturbative nature and reflects 
the "core" dynamics.
To our surprise we find that this is not the case. 
Our numerical data presented in Fig.~\ref{DRtrP}
show beyond any doubt that the "LRT scaling" 
applies within the whole (extended) perturbative regime, 
as in the case of $t_r^E$, 
thus not invoking the perturbation strength $\varepsilon$.
We see that the FOPT expression (\ref{eq:LRT-t_r^P}) 
for $t_{r}^{P}$ shown as a heavy-dashed line describes
the numerical findings.

We conclude that the compensation 
time $t_r$ is mainly related to the dynamics 
of the tails, and hence can be deduced from 
the LRT analysis.

%%%%%%%%%%%%%%%%%%%%%%%%%%%%%%%%%%%
\subsubsection{Decay of $\mathcal{P}(t)$ in the non-perturbative regime}

\begin{figure}[p]
\centerline{\includegraphics[height=0.85\vsize,angle=0,clip]{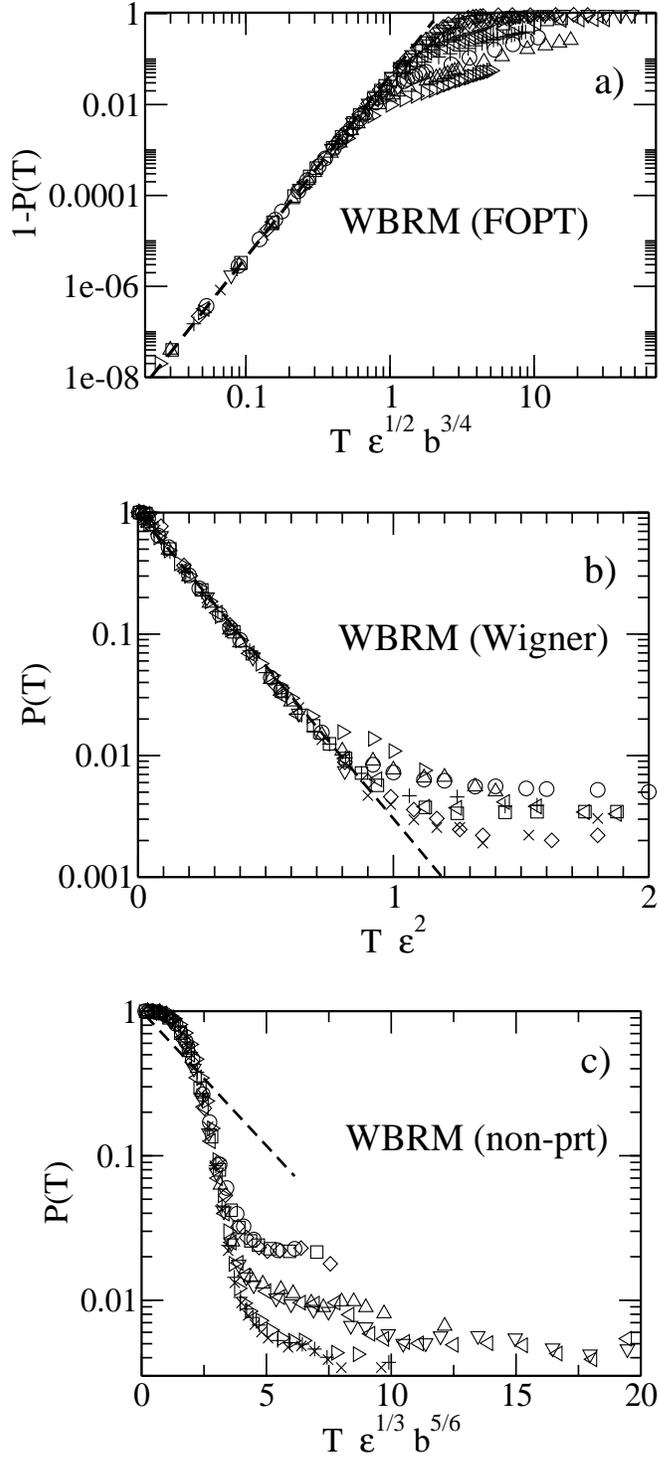}}
\caption{\label{DRPT}
Simulations of driving reversal for the WBRM model. The survival probability $\mathcal{P}
(T)$ at the end of the pulse is plotted against $T$ in the (a) FOPT regime, (b) Wigner 
regime and (c) non-perturbative regime. In (a) the thick dashed line indicates the super-Gaussian
 decay (\ref{eq:superexp}) while (b-c) indicates a Wigner exponential decay 
(\ref{eq:LRT-P(T)-wig}). Various symbols correspond to different ($\epsilon$,b) values 
such that the $\epsilon < \epsilon_c$ in (a); $\epsilon_c <\epsilon < \epsilon_{\rm prt}$
in (b) and $\epsilon > \epsilon_{\rm prt}$ in (c). }
\end{figure}

Let us now turn to the non-perturbative regime 
(see Fig.~\ref{DRPT}c). As in the case of the 
spreading kernel $\delta E(T)$, the decay of ${\cal P}(T)$ 
is no longer captured by perturbation theory. 
Instead, we observe the same non-universal scaling with 
respect to $\varepsilon^{1/3}\times T$ as in the case 
of $\delta E(T)$. 
\be{0}
{\cal P}(T) = h^P_{\tbox{nprt}} \left( b^{5/6} \varepsilon^{1/3}  T  \right)\szp
\ee
The reason is that in the non-perturbative regime 
the two energy scales $\Gamma$ and $\Delta_{b}$,  
which were responsible for the difference 
between ${\cal P}(T)$ and $\delta E(T)$,  
lose their meaning. As a consequence, the spreading process 
involves only one time scale and the behavior of both 
${\cal P}(T)$ and  $\delta E(T)$ becomes similar,  
leading to the same scaling behavior.

%%%%  FIGURE %%%%%%%%%%%%%%%%%%%%%%%%%%
%%%% END FIGURE %%%%%%%%%%%%%%%%%%%%%%%

%%%%  FIGURE %%%%%%%%%%%%%%%%%%%%%%%%%%
%%%% END FIGURE %%%%%%%%%%%%%%%%%%%%%%%

%%%%%%%%%%%%%%%%%%%%%%%%%%%%%%%%%%%%%%%%%%%%%%%%%%%%%%%%%%%%%%%%%%%%%%%
\subsection{Driving Reversal Scenario: 2DW Case\label{sub:DR-2DW}}

In the representative simulations of the 2DW model 
in Fig.~\ref{DRde} (upper left panel) we see that 
the spreading $\delta E(t)$ for $T=0.48$
and various perturbation strengths $\varepsilon$ 
follows the LRT predictions very well.
Fig.~\ref{DRde} (lower left panel) shows that the agreement 
with the LRT is observed for any value 
of the period $T$. This stands in clear 
contrast to the EBRM model shown in Fig.~\ref{DRde} (right panels).

The agreement with LRT in the non-perturbative regime, 
as in the case of wavepacket dynamics, reflects detailed QCC. 
We recall that "to get into the non-perturbative regime"   
and "to make $\hbar$ small" means the same. 
All our simulations are done in a regime where 
LRT can be trusted at the classical ($=$non-perturbative) limit. 
It is only for RMT models that we observe a breakdown 
of LRT in the non-perturbative regime.

%%%%%%%%%%

What about $\mathcal{P}(T)$? This quantity has no classical 
analogue. Therefore QCC considerations are not applicable.
Also LDoS considerations cannot help here. The one-to-one 
correspondence between the LDoS and the survival probability 
applies to the simple wavepacket dynamics scenario 
(constant perturbation).  

It is practically impossible to make a quantitative analysis 
of $\mathcal{P}(T)$ in the case of a real model because 
the band-profile is very structured and there are severe 
numerical limitations. Rather, what we can easily do is 
to compare the 2DW with the corresponding EBRM. Any  
difference between the two constitutes an indication for 
a non-perturbative effect.  Representative simulations  
are presented in Figure \ref{DRpt}.

\begin{figure}[t]
\centerline{\includegraphics[width=0.65\hsize,angle=0,clip]{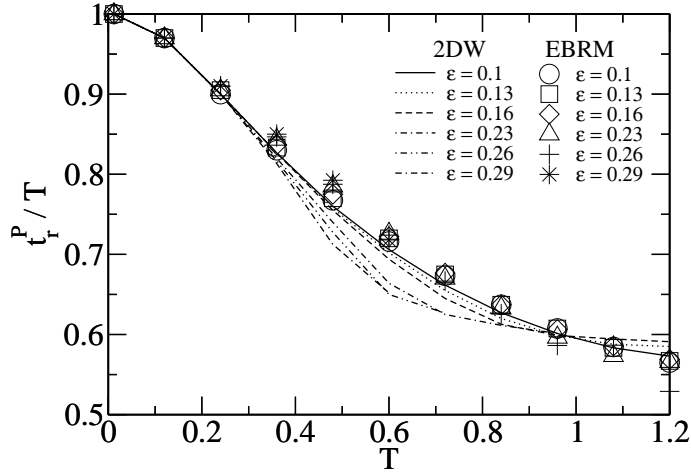}}
\caption{\label{tP}
The compensation time $t_r^P$ 
for 2DW simulations, and for the corresponding 
EBRM simulations as a function of $T$ 
for various values of $\varepsilon$.}
\end{figure}

In Figure \ref{tP} we show the dependence of the compensation time $t_r^P$ on $T$ for 
the EBRM model. We see very nice scaling behavior that indicates that our numerics 
(as far as $\mathcal{P}(T)$ is concerned!) is limited to the perturbative regime. 
We emphasize again that the physics of  
$\mathcal{P}(t)$ is very different from  
the physics of $\delta E(t)$. Therefore, this 
finding by itself should not be regarded as very surprising.
A sharp crossover to a non-perturbative 
behavior can be expected for a ``sharp" 
band-profile only (which is the WBRM and not the EBRM--
see Fig.\ref{DRtrP}).

Now we switch from the EBRM model to the 2DW model. Do we see any deviation from LRT 
scaling? The answer from Fig.~\ref{tP} is clearly yes, as reflected by the $\epsilon$
dependence of the curve. The effect is small, but "it 
is there". It indicates that the ``body'' of the probability distribution, in the case of 
the 2DW dynamics, does not evolve the same way as in the EBRM case. Indeed we know 
that the main part of the distribution evolves faster (in a ballistic fashion rather 
than diffusively), and therefore we observe lower values of $t_r^P$.

\begin{figure}[t]
\centerline{\includegraphics[width=0.62\hsize,angle=0,clip]{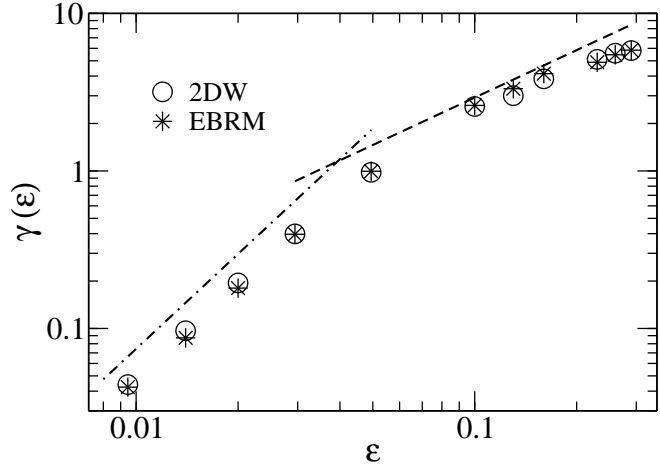}}
\caption{\label{gammaP}
The estimated decay rate $\gamma$ for the same simulations as in the previous figure.  }
\end{figure}

Assuming that the decay of $\mathcal{P}(T)$  
is given by the exponential law, we extract 
the corresponding decay rates $\gamma$. 
It should be clear that the fitting is done merely 
in order to extract a numeric measure for the behavior 
of the decay. We would not like to suggest that the 
decay looks strictly exponential. The results are 
reported in Figure \ref{gammaP}.
We find that for $\varepsilon<\varepsilon_{\tbox{prt}}$, 
the decay rate $\gamma(\varepsilon)\propto\varepsilon^{2}$, 
as expected by Wigner's theory, 
while for $\varepsilon>\varepsilon_{\tbox{prt}}$ 
we find that $\gamma\propto\varepsilon$. 
This linear dependence on $\varepsilon$ is 
essentially the same as in the corresponding 
wavepacket dynamics scenario. There it is 
clearly associated with the width $\delta E_{\tbox{cl}} \propto \varepsilon$ 
of the LDoS.

As far as $\gamma$ is concerned the behavior 
of 2DW and the EBRM models are the same, 
and there is an indication of the crossover 
from the perturbative to the non-perturbative 
regime, as implied (in a non-rigorous fashion)  
by the LDoS theory. It is $t_r^P$ rather than $\gamma$ 
that exhibits sensitivity to the nature of the dynamics. 
This is because $t_r^P$ is sensitive 
to the evolution of the main part of the distribution. 
We already had made this observation on the basis 
of the analysis of the WBRM model (see previous Subsection \ref{sub:DR-WBRM}).  
Here we see another consequence of this observation.

%%%%  FIGURE %%%%%%%%%%%%%%%%%%%%%%%%%%
%%%% END FIGURE %%%%%%%%%%%%%%%%%%%%%%%

%%%%  FIGURE %%%%%%%%%%%%%%%%%%%%%%%%%%
%%%% END FIGURE %%%%%%%%%%%%%%%%%%%%%%%

%%%%%%%%%%%%%%%%%%%%%%%%%%%%%%%%%%%%%%%%%%%%%%%%%%%%%%%%%%%%%%%%%%%%%%%
\section{Conclusions \label{sec:sum}}

There is a hierarchy of challenges in the study of quantum dynamics. 
The simple way to explain this hierarchy is as follows:  
Let us assume that there are two Hamiltonians, ${\cal H}_1$ 
and ${\cal H}_2$, that differ slightly from each other. 
Let us then quantify the difference by a parameter $\varepsilon$. 
Let us distinguish between a FOPT regime, Wigner regime, 
and non-perturbative (semicircle or semiclassical) regime 
according to the line shape of the LDoS. {\em Do we have 
enough information to say something about the dynamics?}

In the conventional wavepacket dynamics, 
one Hamiltonian is used for preparation and 
for measurement, while the other for propagation. 
It is well known that the Fourier transform of 
the LDoS gives the survival amplitude and hence 
$\mathcal{P}(t)$. But what about other features 
of the dynamics. What about the energy spreading $\delta E(t)$ 
for example? It turns out that the answer 
requires more than just knowing the LDoS. 
In particular we observe that in the non-perturbative 
regime physical models differ from the corresponding RMT model. 
In the former case we have ballistic spreading 
while in the latter we have diffusion.

Is there any new ingredient in the study of driving 
reversal dynamics?  {\em Maybe it is just a variation on 
conventional wavepacket dynamics?} The answer turns out to be   
interesting. There is a new ingredient in the analysis.    
This becomes very clear in the RMT analysis 
where we find a new time scale that distinguishes  
between a stage of "reversible diffusion" and a stage  
of "irreversible diffusion". This time scale ($t_{\tbox{sdn}}$) 
can only be probed in a driving reversal experiment. 
It is absent in the study of conventional wavepacket dynamics.

Things become more interesting, and even surprising,   
once we get into details. Let us summarize our main findings. 
We start with the conventional wavepacket dynamics, 
and then turn to the driving reversal scenario.

The main observations regarding wavepacket dynamics 
are summarized by the diagrams in Fig. \ref{brm_diag}.
We always have an initial ballistic-like stage 
which is implied by FOPT. During this stage the 
first order (in-band) tails of the energy distribution 
grow like $t^2$. We call this behavior "ballistic-like" because 
the second moment $\delta E(t)$ grows like $t^2$. 
It is not a genuine ballistic behavior because 
the $r$th moment does not grow like $t^{r}$ 
but rather all the moments of 
this FOPT distribution grow like $t^{2}$.

The bandwidth $\Delta_b$ is resolved at the time $\tau_{\tbox{cl}}$. 
In the perturbative regime this happens before 
the breakdown of perturbation theory, while 
in the non-perturbative regime the breakdown $t_{\tbox{prt}}$ 
happens before $\tau_{\tbox{cl}}$. As a result, in the 
non-perturbative regime we can get a non-trivial 
spreading behavior which turns out to be "ballistic" 
or "diffusive", depending on whether the system 
has a classical limit or is being RMT modeled.

Once we consider a driving reversal scenario, it turns 
out to be important to mark the time $t_{\tbox{sdn}}$ when the 
energy distribution is resolved. The question is ill-defined 
in the perturbative regime because there the energy distribution 
is characterized by two energy scales (the "bandwidth" and the 
much smaller "core width"). But the question is 
well-defined in the non-perturbative regime where 
the distribution is characterized by one energy scale. 
It is not difficult to realize that for ballistic 
behavior $t_{\tbox{sdn}}\sim\tau_{\tbox{cl}}$ which is also the 
classical ergodic time. But for diffusion we get 
separation of time scales   $t_{\tbox{prt}} \ll t_{\tbox{sdn}} \ll \tau_{\tbox{cl}}$. 
Thus we conclude that the diffusion has two stages: 
One is reversible while the other is irreversible.

But the second moment does not fully characterize the 
dynamics. In the other extreme we have the survival probability. 
Whereas $\delta E(t)$ is dominated by the tails, $\mathcal{P}(t)$
is dominated by the "core" of the distribution. Therefore 
it becomes essential to distinguish between the FOPT regime 
where the "core" is just one level, and the rest of the 
perturbative regime (the "Wigner" regime) where 
the core is large (but still smaller compared with the bandwidth).

The main findings regarding the driving reversal 
scenario are summarized by the following table:

%%%%%%%%%%%%%%%%%%%%%%%%%%%%%%%%%%%%%%%%%%%%%%%%%%%%%%%%%%%%% 

\ \\ \ \\ 
{\scriptsize
\begin{tabular}{|l|c|l|l|l|}
\hline

Regime &
perturbation  strength&
$\mathcal{P}(T)$ behavior &
$t_r$ behavior &
$\delta E(T)$ behavior \\

\hline

1st order perturbative  &
$\varepsilon < \varepsilon_c$ &
LRT &
LRT &
LRT  \\

\ &
\ &
(super-Gaussian) &
\ &
(ballistic-like) \\

\hline

extended perturbative&
$\varepsilon_c  < \varepsilon < \varepsilon_{\tbox{prt}}$ &
Wigner &
LRT(!) &
LRT \\

("Wigner") &
\ &
(Exponential) &
\ &
(ballistic-like) \\

\hline

non-perturbative &
$\varepsilon > \varepsilon_{\tbox{prt}}$ &
non-perturbative   &
non-perturbative   &
non-perturbative$^*$  \\

\ &
\ &
(non-universal) &
(non-universal) &
(diffusive/ballistic) \\

\hline
\multicolumn{5}{|l|}{
$^*$for the WBRM we have diffusion while for the 2DW model we have ballistic
 behaviour as implied by classical LRT}\\

\hline

\end{tabular}
}
\ \\ \ \\

%%%%%%%%%%%%%%%%%%%%%%%%%%%%%%%%%%%%%%%%%%%%%%%%%%%%%%%%%%%%%%%%%%%%%%%%%%%%

As expected we find that $\mathcal{P}(T)$ obeys FOPT behavior 
in the FOPT regime, which turns out to be super-Gaussian decay.
In the Wigner regime $\delta E(T)$ still obeys LRT because 
the tails obey FOPT, while the non-perturbative core 
barely affects the second moment. But in contrast to that $\mathcal{P}(T)$ is 
sensitive to the core, and therefore we find Wigner (exponential) 
decay rather than FOPT (super-Gaussian) behavior. 
However, when we look more carefully at the whole $\mathcal{P}(t)$ 
curve, we find that this is not the whole story. 
We can characterize  $\mathcal{P}(t)$  by the compensation time $t_r$. 
It turns out that $t_r$ is sensitive to the nature of the dynamics. 
Consequently it obeys "LRT scaling" rather than "Wigner scaling". 
This has further consequences that are related 
to quantal-classical correspondence. Just by looking 
at $\mathcal{P}(T)$  we cannot tell whether we look at
the "real simulation" or on its RMT modeling. But looking 
on $t_r$ we can find a difference. It turns out that in the 
physical model $t_r$ exhibits $\varepsilon$ dependence, 
while in the case of RMT modeling $t_r$ is independent of $\varepsilon$
and exhibits "Wigner scaling".

Finally we come to the non-perturbative regime. Here we have, 
in a sense a simpler situation. We have only one energy scale, 
and hence only one time scale, and therefore  $\delta E(T)$
and $\mathcal{P}(t)$ essentially obeys the same scaling. 
Indeed we have verified that the non-perturbative scaling 
with  $t_{\tbox{sdn}}$ in WBRM simulations is valid for both the second 
moment and the survival probability.

Finally we would like to emphasize that the notion 
of  "non-perturbative" behavior should not be confused 
with "non-linear"  response. In case of quantized models, 
linear response of the energy spreading $\delta E(T)$
is in fact a consequence of non-perturbative behavior.   
This should be contrasted with the WBRM model, 
where QCC does not apply, and indeed deviations 
from the linear response appear once we enter the non-perturbative 
regime.

The study of irreversibility in a simple driving reversal scenario 
is an important step towards the understanding of irreversibility 
and dissipation in general. The analysis of dissipation reduces  
to the study of energy spreading for time dependent Hamiltonians ${\cal H}(Q,P;x(t))$. 
In generic circumstances the rate of energy absorption is determined 
by a diffusion-dissipation relation: The long time process of dissipation 
is determined by the short time diffusion process. The latter is related 
to the fluctuations $\tilde{C}(\omega)$ via what we call "LRT formula". 
Thus the understanding of short time dynamics is the crucial step   
in establishing the validity of the fluctuation-dissipation relation.

%%%%%%%%%%%%%%%%%%%%%%%%%%%%%%%%%%%%%%%%%%%%%%%%%%%%%%%%%%%%%%%%%%%%%%%%%%%%%
%%%%%%%%%%%%%%%%%%%%%%%%%%%%%%%%%%%%%%%%%%%%%%%%%%%%%%%%%%%%%%%%%%%%%%%%%%%

\ \\

\section*{Acknowledgments}
This research was supported by a grant from the GIF, the German-Israeli
Foundation for Scientific Research and Development, 
and by the Israel Science Foundation (grant No.11/02). 

%\bibliographystyle{elsart-num}
%\bibliography{paper,books}

\end{document}